# Oprema –
# The Relay Computer of Carl Zeiss Jena

Jürgen F. H. Winkler

*Friedrich Schiller University, Jena*

The Oprema (**Op**tik**re**chen**ma**schine = computer for optical calculations) was a relay computer whose development was initiated by Herbert Kortum and which was designed and built by a team under the leadership of Wilhelm Kämmerer at Carl Zeiss Jena (CZJ) in 1954 and 1955. Basic experiments, design and construction of machine-1 were all done, partly concurrently, in the remarkably short time of about 14 months. Shortly after the electronic G 2 of Heinz Billing in Göttingen it was the 7[th] universal computer in Germany and the 1[st] in the GDR. The Oprema consisted of two identical machines. One machine consisted of about 8,300 relays, 45,000 selenium rectifiers and 250 km cable.

The main reason for the construction of the Oprema was the computational needs of CZJ, which was the leading company for optics and precision mechanics in the GDR. During its lifetime (1955–1963) the Oprema was applied by CZJ and a number of other institutes and companies in the GDR.

The paper presents new details of the Oprema project and of the arithmetic operations implemented in the Oprema. Additionally, it covers briefly the lives of the two protagonists, W. Kämmerer and H. Kortum, and draws some comparisons with other early projects, namely Colossus, ASCC/Mark 1 and ENIAC. Finally, it discusses the question, whether Kortum is a German computer pioneer.

## Pre-history

At the beginning of the 20[th] century, three people were born, who were to play an important role in the history of computing in Germany. Wilhelm Kämmerer was born on 23 July 1905 in Büdingen, a small town north-east of Frankfurt/Main; Herbert Franz Kortum was born on 15 September 1907 in Gelting, near Schleswig in the northern part of Germany, and Konrad Ernst Otto Zuse was born on 22 June 1910 in Berlin.

Zuse was not directly involved in the activities around the Oprema, but his person can serve as a kind of reference point in the history of computing in Germany, as on 12 May 1941 he presented the first program-controlled computer, the Z3, in Berlin to visitors from the German Aeronautics Research Institute.[1] Moreover, indirect connections between the Oprema, Kämmerer, Kortum and Zuse existed during the development period and later after the reunification of Germany.

Kämmerer studied mathematics at Gießen and Göttingen, obtained a PhD in mathematics from the university of Gießen in 1927, and, in 1929, became a high school teacher of mathematics in Naumburg/Saale, in the central part of Germany.[2] In 1943 Kämmerer joined the Carl Zeiss Company in Jena (CZJ) and became a member of Kortum's development lab[5]. At the end of the war he returned to Naumburg/Saale and took up work again as a high school teacher for about one year. On 1 June 1946 he joined CZJ again.[4]

Kortum studied physics at today's Friedrich Schiller University (FSU) in Jena from 1926 to 1930, obtained a PhD in Physics in 1930 and then worked as an assistant professor at the Institute of Physics at the same university.[5] In 1934 Kortum left the university and joined CZJ, which was one of the leading optical companies worldwide. His first project was "Entwicklungsarbeiten an Rechengeräten für die Feuerleitung unter Verwendung elektromechanischer Analogrechenglieder und Servosysteme [Development of computational devices for fire control using analogue computing elements and servo systems]".[6] In 1939 Kortum became the co-head of the reorganized Telegroup, the development lab for field glasses and other optical equipment for military use[7], which was then called KoKor (Konstruktionsbüro Kortum [Development Lab Kortum]).[8] This lab also developed sophisticated electro-mechanical targeting devices for Luftwaffe bombers including the Lotfe 7D bombsight, which Kurowski calls "a miracle of technology", and the TSA (Tief- und Sturzanlage [level- and dive-bombing aiming device]). Kortum's work also became known in the USA. On 13 April 1945 US troops entered Jena and also took command at CZJ. The US specialists were especially interested in the aerial cameras and the aiming devices. Kortum was interrogated by Captain James Harris about the TSA in late April or early May 1945. It is interesting to note that the Germans learned about dive-bombing from the Americans when Ernst Udet, who was a notable flying ace of World War I, visited the National Air Races in Los Angeles in 1928 and later saw the Curtiss Hawk F8C ("Helldiver") at the Curtiss airfield on 27 September 1933.[9] When the US military left CZJ and Jena at the end of June 1945, they took, amongst other people and things, Kortum and his family with them and brought them to Heidenheim in southern Germany.[10] In October 1945 he was still imprisoned there.[11]

In Jena the Soviets took command at CZJ on 1 July 1945. Karl Schumann became the new head of KoKor.[12] The Soviets were also very eager to learn about the devices that had been developed by KoKor. In a memo Schumann reports that Lieutenant Colonel Urmajeff was very interested in a high-precision gyroscope, a prototype of which had been built in 1940 on Kortum's order. Urmajeff remarked that such devices would be quite important for Russian aircrafts when they had to fly over the vast lands of northern Siberia or over the North Pole to America.[13]

At the beginning of 1946, Kortum returned to Jena without his family, which was staying in Königsbronn near Heidenheim. The exact date of his return is somewhat unclear, different dates are mentioned in different sources. The most precise primary source is a memo of Schumann dated 12 February 1946. In this memo Schumann reports that Major Gleinick and First Lieutenant Chorol had visited him on that day. They had said that they would come again on 13 February and had asked that Dr. Kortum should then explain the complete structure of the TSA. These circumstances suggest that Kortum was back in Jena not later than 12 February 1946.[14]

During 1945 and 1946 the Soviets were quite interested in the bombsights, that had been developed by KoKor, and wondered how they could make use of them for their own aircrafts. In August and September 1946 e.g., Kortum did experiments near Jüterbog, south of Berlin, with aiming devices mounted on Soviet army airplanes.[15]

## History
How the idea for the Oprema (Optikrechenmaschine = computer for optical calculations) evolved and who came up with the idea first is unknown. None of those involved, who are now all dead, has left any description of this process. The first primary source that mentions the Oprema is a memo by Schumann dated 2 Nov. 1945.

Schumann reports that Lieutenant Colonel Urmajeff had visited him on that day and had asked about the Oprema. Schumann had answered that "Oprema" was currently more an idea and by far not a mature project. Urmajeff asked again on 6 Nov. 1945 and Schumann gave exact the same answer, that "Oprema" was rather an idea than a project.[16]

The next hint on the Oprema occurs in the minutes of a meeting on 17 Aug. 1946 attended by Major Turügin, Major Jachantoff, Dr. Tiedeken, Schumann and Dr. Kortum. Kortum made a short statement about the plans for the development of a computer for optical calculations based on the components that had been developed by EBo 5 (Entwicklungsbüro 5 [development lab no. 5]). Since July 1945 "EBo 5" was the name of Kortum's former development lab "KoKor". The components were those of the bombsights and other devices that had been developed there during the preceding years. These components were elements of analogue computation techniques. Kortum said that he could not give any details because this was a project still in its infancy, and that a lot of theoretical work had to be done before a design could begin. First of all, he needed a scientist of suitable background and experience. Major Turügin asked for a more detailed written report explaining how Kortum thought this project would proceed. This report was to be delivered by 21 Aug. 1946. In a later meeting with Major General Nikolaew it was decided that Kortum's report would be handed in later because at that time Kortum was apparently very busy and overwhelmed with work (see e.g. the experiments near Jüterbog mentioned above).[17]

Later in the meeting with Major Turügin on 17 Aug. 1946, Dr. Tiedeken presented his idea to use Hollerith machines for computations required for the design of optical systems. He said that he had already contacted the Hollerith company and had received a positive response. The "Hollerith company" must have been the DEHOMAG (Deutsche Hollerith-Maschinen Gesellschaft mbH [German Hollerith Machines Company Ltd.]) that was founded on behalf of Herman Hollerith on 30 Nov. 1910 and that had restarted the production of punched-card machines in Berlin in Oct. 1945.[18] Dr. Tiedeken was a scientist in the lab for the calculations for photographic lenses.[19]

But these plans and ambitions came to a sudden halt on 22 Oct. 1946. In the early morning of that Tuesday the Operation Osoaviakhim was put into action. All over the Soviet Occupation Zone Soviet officers, accompanied by a group of soldiers, were knocking on the doors of several thousand scientists, engineers, technicians and skilled workers employed in war-related industries. The officer read out an order of the Soviet Military Administration that the resp. person had to do professional work in the Soviet Union (SU) under comparable conditions as their Soviet counterparts for 5 years, and that he could take with him his wife, children and as much of his belongings as he wanted.[20] In Jena 276 personnel from Zeiss alone were thus deported to different locations in the SU, among them Kämmerer, with his wife and two daughters, and Kortum, without his family which was still in the American Occupation Zone. Kämmerer even took his Steinway Concert Grand with him (and in 1953 back) and Kortum his violin. In the following years, the group around Kortum and Kämmerer (K&K) lived at several places mostly near Moscow: Mamontowka (Nov. 1946 – Nov. 1948), Moscow-Sokolniki (Nov. 1948 – Jan. 1952), Krasnogorsk (Jan. 1952 – Jun. 1952), Gorodomlia (Jun. 1952 – Nov. 1953). Instead of 5 years they had to stay 7 years in the SU.[21]

K&K gave no details about the kind of work they had to do in those years. It may be assumed that, before they could return to Germany, they had to sign a formal declaration of obligation not to tell anybody about their work in the SU, similarly as the rocket scientist Kurt Magnus had to do. After the German Reunification Kämmerer wrote: "Unsere Tätigkeit bestand im wesentlichen darin, unsere Arbeitsmethoden und Erfahrungen an sowjetische Ingenieure zu übertragen [Our work essentially was to transfer our methods and experience to Soviet engineers]". Other sources say that the Kortum group worked on bombsights as it had done in the preceding years at CZJ. This would be analogous to what the rocket scientists Werner Albring and Magnus or the physicist Kurt Berner reported about their stay as "specialists" in the SU. To my knowledge none of the Zeiss people has published anything similar to the reports of Albring, Berner and Magnus. A similar observation has been made independently by Matthias Uhl.[22]

The detention on the small island of Gorodomlia in Lake Seliger, halfway between Moscow and Leningrad (now St. Petersburg again), was, by the Soviet authorities, intended to be a cooling-off period, during which the scientists should "forget" the details of the work they had done before. Kortum, Kämmerer and the other Zeiss scientists were completely isolated on this small island. There was no scientific or even belletristic literature. The only books they had were those they had brought with them in October 1946. When they left Krasnogorsk they had to leave behind all papers, notes, even letters they had received from relatives in Germany; it all was burned.[23]

Helga Kämmerer reports that during their stay on Gorodomlia K&K, who had no official work to do, discussed their ideas for the Oprema. But unfortunately no direct statement by them seems to exist on these discussions. In a report dated May 1958, Kortum states:

"Als dann aber tatsächlich der Plan realisiert wurde und einige Tage vor Jahresende die beiden Maschinen mit Kosten, die mit den geplanten Kosten in guter Übereinstimmung waren, fertig montiert vorgestellt werden konnten, wurden [wurde; JW] von vielen Seiten die Version aufge-bracht, wir hätten diese Entwicklung bereits fertig in der Tasche aus der SU mitgebracht. Dazu sei festgestellt, daß dies eine freie Erfindung der Urheber dieser Version ist, insofern, als wir uns hinsichtlich dieses Themas während unseres Aufenthaltes in der SU zwar durch Literaturstudien laufend über die in der Welt vorgegangene Entwicklung informieren konnten und insofern allerdings die nötigen Fachkenntnisse und konkrete Vorstellungen zur Lösung der gestellten Aufgabe mitbrachten, wir hatten aber weder Zeit noch Gelegenheit, konkrete Vorarbeiten dazu durchzuführen und mitzunehmen. Es ist daher die Wahrheit, daß Entwicklung, Bau und Montage der gesamten Anlage mit allen dazu nötigen Arbeiten im Jahre 1954 durchgeführt worden sind.

[When the plan [to build the Oprema; JW] had really been realized and, several days before the end of the year, the two machines could be presented as completely assembled, and the costs were as expected, a lot of people spread the rumor that we had brought the complete plans of the Oprema with us when we returned from the SU. This, I must say, is a free invention by the initiators of this rumor. It is true that during our stay in the SU we could indeed study the literature and thus follow the worldwide developments. Therefore, we came back with the know-how and the concrete ideas necessary for the solution of this task. But we neither had the time nor the opportunity to do any specific preliminary works, so we could have brought the results back with us. It is therefore the truth that the development and construction of the whole system and all the activities necessary for them were done in 1954]."[24]

Helga Kämmerer also reports that all the knowledge of K&K about western computers came from broadcasts of the BBC, which came in rather weakly.[25] This does not correspond to Kortum's statement mentioned above. She gives no further details about these broadcasts, e.g. the time or to which specific World Service of the BBC (e.g. German, English, Russian) they had listened. It seems that we can rule out that K&K heard the five broadcasts on "Automatic Calculating Machines" which were broadcast on BBC's Third Programme radio service in May and June 1951. The Third Programme, started on 29 September 1946, was a home service, which was barely received in the whole country. In April 1951 "The Third's coverage was now estimated to have increased to between sixty-five and seventy percent of the population, . . ."[26]

The Zeiss people on Gorodomlia were allowed to leave the SU in November 1953 and arrived in Jena on 21 November 1953, where they were received by their Zeiss colleagues in the hall of mirrors of the hotel "Schwarzer Bär [Black Bear]". On 22 November they joined CZJ again. Kortum then took a convalescent leave and started work as Chief of Development on 21 January 1954.[27]

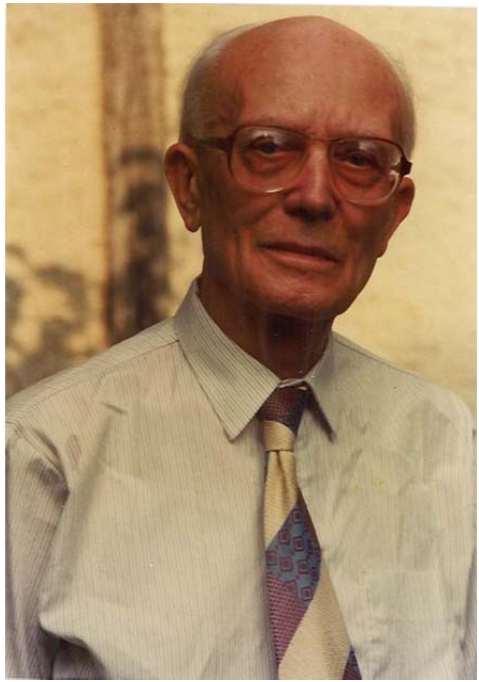

**Fig. 1. Wilhelm Kämmerer.**
**Courtesy of Helga Kämmerer.**

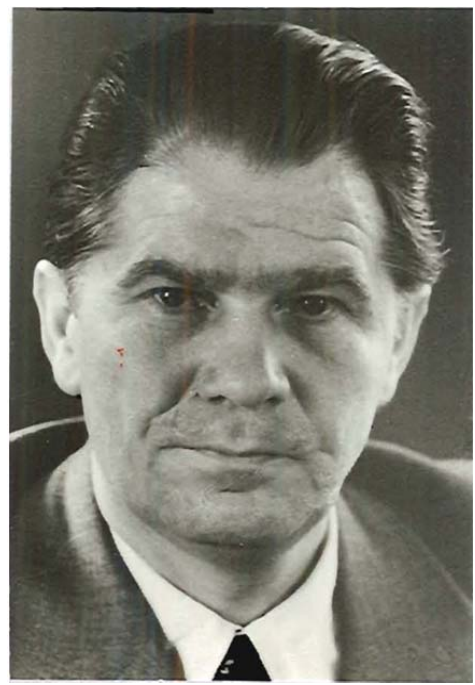

**Fig. 2. Herbert Franz Kortum.**
**Courtesy of Helga Kämmerer.**

## Design and Construction of the Oprema

During the winter 1953/54, K&K visited Nikolaus Joachim Lehmann at the Dresden Institute of Technology. Lehmann, a young physicist, had started a small project to develop an electronic computer after he had read the paper on ENIAC in MTAC July 1946, which had reached him only in early spring 1948! No details of this meeting are reported, neither in the papers of K&K nor in Lehmann's papers in the archives of the Deutsche Museum in Munich. Since Lehmann's computer was expected to be ready only in late 1955, K&K started their own computer project at CZJ.[28]

The first document of this phase of the Oprema story is the first entry in Gerhard Lenski's work diary dated 5 April 1954. This handwritten diary does not bear his name, but his role as head of physical design and construction, his handwriting, the attribution by Edgar Mühlhausen in his paper "Am Anfang war OPREMA . . . [In the Beginning was the OPREMA . . .]", and, last but not least, the very contents of this document make the authorship of Lenski quite certain. Mühlhausen joined the Oprema project on 6 September 1954, right after he had finished his studies of mechanical and electrical engineering at the Dresden Institute of Technology. Lenski's work diary (LWD) covers the time from 5 April to 24 December 1954 and is 58 pages long. It contains entries for all working days of that period, i.e. all days except Sundays and holidays because at that time Saturday was a regular work day in the GDR. The only Saturday without an entry is the Saturday before Easter.

LWD begins as follows: "

1954.        Arbeitsberichte
5.4.   Arbeitsbeginn. Allgemeine Einführung und Auftrag zur Konstruktion der „Rechenmaschine für optische Rechnungen." Bespr. bei Dr. Kämmerer. Relaismaschine Zu Kontrollzwecken als Zwillingsmaschine ausgeführt. Rechnet im Dualsystem. Aufbau des Gerätes in grossen Zügen.

Dualsystem mit Aiken Verschlüsselung. . . .

[1954.        Work Reports
5.4.   Work begins. General introduction and assignment for the design of the "Calculating Machine for Optical Calculations." Meetg. at Dr. Kämmerer's. Relay machine For checking purposes conceived as twin-machine. Computes with the binary system. Broad outline of the device. Binary system using the Aiken code. . . .]"[29]

During the first five weeks there were almost daily meetings at Kämmerer's office. The LWD shows that Kämmerer already had quite detailed ideas about the computer. The computer was to be based on relays and to use a BCD (binary coded decimal) number representation. The program and input numbers were to be be stored in plugboards. The plugboards were to be permanently installed. Each instruction and input number was to occupy one row of the resp. plugboard.  The loading of a program and the input numbers had to be done by putting plugs into certain sockets, where each such plug-socket pair corresponded to one bit. Lenski proposed to use switches instead of the plug-socket pairs. Kämmerer refused because he wanted to use plugging templates in order to facilitate the loading of programs, which is a rather error-prone process. Kämmerer also had detailed ideas about the logical structure, the principles of control, the instruction format, and the number format.

The instruction format was as follows:

Adr1(6)  Op(6)  Adr2(6)  Adr3(5)  Adr4(4)

where the numbers in parentheses are the number of bits of the corresponding part of the instruction. One instruction therefore consisted of 27 bits. More details about the instruction code are given below in the section on the technical properties of the Oprema.

Numbers were to be stored as floating-point numbers in binary coded decimal form in the following format:

S DDDDDDDD S E

where

- S is a sign (+ or −, 1 bit),
- D is a binary coded decimal digit of the significand (4 bit),
- E is the binary coded exponent (4 bit).

Kämmerer calls the significand the "Mantisse [mantissa]" of the floating point number, whereas I follow the IEEE and ISO standards and use the term "significand". Donald E. Knuth criticized the use of the term "mantissa" in this context: "but it is an abuse of terminology to call the fraction part a mantissa, since this concept has quite a different meaning in connection with logarithms." Interestingly, in the paper "Oprema, die programmgesteuerte Zwillingsrechenanlage … [Oprema, the program-controlled twin-computer …]" the term mantissa is always written in quotation marks as „Mantisse". It seems, that Kämmerer as a mathematician was aware of the problem mentioned by Knuth.

As mentioned above, the Aiken code was to be used for the digits of the significand, but on Tuesday 24 April 1954 it was decided to use the excess-3 code instead, because this led to a lower load of the relay contacts.

On the quantitative side Kämmerer's plans for one machine envisaged:

- 5 plugboards with a total of 200 rows à 40 sockets for the input numbers,
- approx. 6,000 relays and 20,000 rectifiers,
- 442 lamps, 300 push-buttons and 300 switches at the console. (LWD 6 and 8 April)

LWD contains no estimation for the number of program plugboards and their rows.

The final figures for one machine were:

- 6 plugboards with a total of 300 rows à 27 sockets for the program,
- 1 plugboard with 28 rows à 39 sockets for input numbers,
- 4 plugboards with a total of 320 rows à 41 sockets for input numbers,
- 8313 relays and approx. 45,000 rectifiers.

The available documents contain no figures on the final numbers of lamps and switches at the console.[30]

Kämmerer had already done some significant design work before April 1954, especially after his return to Jena on 21 November 1953.[31] Whether any important design work had been done before that date back in the Soviet Union is currently unclear and has already been discussed above in the context of Kortum's report of May 1958.

In April 1954 work on the Oprema started on a rather small scale involving only a handful of persons. Nevertheless, by mid-May the overall design and respective designs for the adder, the control pyramid, the control matrix, the command chain, the instruction processing and for the input and output process had been drawn up. Since all arithmetic operations were to be based on the addition operation the adder was the central part of the arithmetic unit. The output was to be printed on a modified electrical typewriter. Additionally, a plan of the physical layout, and, especially for a meeting in Berlin, a schedule and a cost estimate had been prepared.

The next important event in the history of the Oprema was a meeting at the Ministerium für Maschinenbau [Ministry of Machine Building] in East Berlin on 17 Mai 1954. Minister Heinrich Rau had summoned Dr. Hugo Schrade, the plant manager of CZJ, and Dr. Kortum, the Head of Development. Rau first demanded that CZJ should produce 100 exemplars of the newly developed photocolposcope model in due time for the Leipzig Trade Fair. Additionally, Rau expressed his worries about the insufficient plan fulfillment at CZJ. After the discussion of some further problems CZJ had with the authorities, Kortum presented his proposal, based on plans prepared during the preceding months, for building the Oprema, which would be able to do the work of at least 120 human computers. As motivation he mentioned that large-scale computers had already been in use (in some cases for several years) in other countries (America, Sweden, Switzerland, the UK). He pointed out that some rival firms had already been using automatic computers for several years and gave two special examples:

- the Leitz company in Wetzlar, well-known for the Leica camera, had recently installed the Zuse Z5 computer;
- the Swiss company Wild at Heerbrugg had recently presented a new lens for aerial cameras, the Aviogon, which had been designed by Ludwig Bertele, who might have used the "computer installed in Zurich" for the necessary computations.[32]

Leitz had ordered the Z5, a relay computer, from Konrad Zuse in 1950 and it was delivered to Wetzlar on 7 July 1953.

Wild had presented the Aviogon at the 7th International Symposium on Photogrammetry 1952 in Washington, and it had impressed the specialists of CZJ very much. The developer of the Aviogon was Ludwig Bertele, whom the Zeiss people knew quite well because he had worked for Zeiss Ikon in Jena and Dresden between 1926 and 1942. He was one of the most outstanding lens designers in the world. Kortum said that he assumed Bertele had used the "machine installed in Zürich" for the development of the Aviogon. The "machine installed in Zürich" is obviously Zuse's Z4. An attempt to verify Kortum's conjecture produced a negative result; no information on any use of the Z4 by Bertele could be found.[33]

Kortum's presentation seems to have convinced minister Rau and he immediately allotted 1 Mio. DM[34] from the fund for rationalization to this project. In return, Kortum promised to complete the machine by the end of 1954. This meeting took a similar course as a meeting at the Ballistic Research Laboratory (BRL) at Aberdeen Proving Ground on 9 April 1943: John G. Brainerd, John Mauchly and J. Presper Eckert of the Moore School of the University of Pennsylvania had visited Colonel Leslie Simon, the director of BRL, and had presented the ENIAC proposal. "The same day, Simon, convinced that the bold project was of extreme importance, gave his support and promised to include the project in his budget."[35]

On 18 May 1954, the day after the meeting in Berlin, Lenski writes in LWD: "Bespr. Kortum. Mitteilung dass Bau der Maschine genehmigt ist. Sofort mit Arbeit im Grossen anfangen. [Meetg. Dr. Kortum. Announcement that construction of the machine has been approved. Immediately going to begin with the work on a large scale.]".

During the following months the workforce working on the Oprema grew steadily and reached its highest number on 17 December, when e.g. 2 shifts of about 7 persons each worked on the cable harnesses, alone. The last but one cable harness was completed on 17 December. Eberhard Dietzsch, a former lens designer at CZJ, who had joined CZJ as a computer in 1954, recalls that he and his colleagues were released from their computational duties in December and then also worked on the cable harnesses during the graveyard shift.[36] Beginning on 13 December soldering was also done in two shifts. For several weeks before 17 December, 8 switchboard people and even 10 people from the payroll office worked on the Oprema from 15:00 to 22:00, after their regular daily working hours. Lenski gives no exact figures on the workforce involved, but the number of people invited to the Oprema Party on the evening of Saturday, 8 January 1955, was about 260. This figure includes some high-level managers and also people which did not work directly on the Oprema but provided material and component parts. On three lists, 80 persons are mentioned eligible to different kinds of bonuses.[37]

During 1954 not only the work force but also the daily working hours grew steadily. On Thursday, 5 August, Lenski begins to record the time he starts work in the morning: "Anf. 7.$^{30}$ [Beg. 7.$^{30}$]". From 6 August to 28 September he begins at 07:00 and from 29 September to 20 October at 06:30. On 21 October he returns to 07:00 and from 13 to 24 December he begins at 06:30 again. On Tuesday,

28 September 1954, Lenski begins to record the end of his daily work, too: "7 $^{00}$ – 18 $^{30}$". From 11 to 20 November it is mostly 07:00 to 19:00, from Monday 22 November to 10 December 07:00 to 22:00, and, finally, from 13 to 24 December 06:30 to 23:30. These figures differ somewhat from a statement by Lenski made in a TV interview: "Ab Ende Mai Tag und Nacht gearbeitet [From the end of May we worked day and night]". On a list, in which Lenski recorded the overtime done by different people, he states 373.5 hours of overtime during 10 months for himself.[38]

The last entry in LWD reads: "
20.12– 24.12   An der Maschine schliesst sich ein 6 $^{30}$ – 23$^{30}$   Loch nach dem andern Relaisplatten sind alle eingebaut. Steckertafeln fehlen nur noch ungefähr. 17 Stück, die nun aber auch schon fertig sind. Der Impulsgeber läuft ohne Prellungen. Er ist in 3 Nächten hinjustiert worden. Das Schaltpult ist bis auf eine Platte auch fertig. Die Ladegleichrichter und Quecksilberrelais sind eingetroffen und werden eingebaut. [
20.12– 24.12   At the machine gap after gap is 6 $^{30}$ – 23$^{30}$   being closed All relay plates have been mounted. Only approx. 17 plugboards are missing, but they have already completed now. The pulse generator runs without bounces. It has been adjusted during 3 nights. Except for one plate, the console is also complete. The charging rectifiers and the mercury relays have arrived and they are now being installed.]"

On Thursday, 30 December 1954, one day before the deadline, CZJ reported the completion of the Oprema to minister Rau. Kämmerer himself later stated that at the end of 1954 the Oprema was "im Bau vollendet [construction was completed]", and that during the debugging of machine-2 in 1955 they found 452 faults, among them 123 missing wires or rectifiers and 47 cold solder joints.[39]

On Wednesday, 5 Jan. 1955, the Under Secretary in the Department for Machine Building, Helmut Wunderlich, visited CZJ, and, as part of this visit, K&K proudly showed him the Oprema. On the pictures taken at this event, the Oprema looks quite complete, at least as seen from the outside. The end of the main construction phase was celebrated with a party at the trade union building on the evening of Saturday, 8 Jan. 1955. Approximately 250 people attended. Someone even composed a poem for this event:

"Oprema – Ballade !

Durch die Hallen von Oprema,
ziehn die Hirten morgens an ihren Platz
und sie schuften an den Platten
als wenn's ging um einen großen Schatz.
Selbst die Herren Konstrukteure
wissen auch bald nicht mehr ein noch aus
sie zerbrechen sich die Köpfe,
wenn es sein muß noch zu Haus.

Refrain:
Tralla, tralla
Prima, prima, beinah wie Oprema,
das baut man in Jena
trallalala.

Oprema hat es in sich,
das weiß ein jeder hier,
und wenn das nicht bald aufhört,
dann wird die Bude leer.

Refrain:
Doch muß man noch dazu sagen,
dass es hier auch schwere Fälle gibt,
und ein jeder wird sich fragen,
wer wird hier zuerst verrückt ?
Ja, man kann hier viel erleben,
alle freun sich nur noch auf eins,
eine Prämie wird es geben,
nur für manche des Vereins.

Refrain:

( Melodie: Bravo, bravo )

[          The Oprema – Ballad !

Through the halls of Oprema
the shepherds move to their place in the morning
and they graft on the panels
as if it was about a precious treasure.
Even the gentlemen designers
are soon at their wits' end
they rack their brains
if necessary even back at home.

Refrain:
Tralla, tralla
Super, super, almost like Oprema,
that is being built in Jena
trallalala.

Oprema is a class of its own,
that is known by everyone,
and if that does not stop rather soon
the hut will empty out.

Refrain:
But there must also be said
that there are serious cases here, too,
and everyone will ask themselves
who will be going crazy first ?
Yes, one may witness here a lot,
all are looking forward to one thing,
there will be a bonus
only for some of the club.

Refrain:

( Tune: Bravo, bravo ) ]

The line "who will be going crazy first ?" sounds like an echo of Brainerd's remark in 1946: "mental breakdowns lurking around each weekend".

Brainerd was the supervisor in responsible charge of the ENIAC project at the Moore School of UPenn.

In 1999, Mühlhausen, who joined the Oprema project in autumn 1954, mentions the following people, who made important contributions to the creation of Oprema: Hans Dietrich, Alfred Jung, Gerhard Lenski, Wilhelm Pöll, Fritz Straube and the engineer Knothe.[40]

During the first months of 1955 Oprema-1 was thoroughly checked and faults, as, for example, cold solder joints or missing wires, were eliminated. In May the first trial runs were carried out, and on Monday, 1 August 1955, the computing center of CZJ was founded and the productive operation of Oprema-1 began. At the beginning of 1956 Oprema-2 was put into productive use, too.[41]

Given these dates, we may estimate how long it took to create the Oprema, though we have to keep in mind that the Oprema was two separate computers. The official figure given has always been 7½ months, which corresponds to the time span between 17 May 1954, the day of minister Rau's approval of the Oprema project, and 30 December 1954, when the "completion" was reported to minister Rau.[42] But, as the LWD show, work began earlier, and machine-1 was first ready for use between May and August 1955. Machine-2 was first ready for use at the beginning of 1956. The LWD begin at 5 April 1954 and for machine-1 we may assume mid-June 1955 as the date it was first ready for use. Unfortunately, I did not find any record stating when Oprema-1 executed successfully its first program, such as the famous log entry of 6 May 1949 for EDSAC: "Machine in operation for first time."[43] For Oprema-1 we thus obtain $14^1/_3$ months (5 April 1954 to 15 June 1955) as a rough estimate of the time for design and construction, which is still a remarkably short period compared to, e.g. the ASCC/Mark I (47 months) or the ENIAC (30½ months). An even shorter time is reported for Colossus-1: "the Dollis Hill engineers, having decided to make an all-electronic processor, took only 11 months to get the first machine into service".[44] As we have seen above in the discussion of the time needed for Oprema-1, there is no clear-cut method to determine such time periods. The 47 months needed for ASCC/Mark I represent the time between the approval of the project by IBM president Thomas J. Watson in February 1939 and the first successful execution of a test program in Jan 1943.[45] With regard to ENIAC Brian Randell says that the memo of John W. Mauchly of August 1942 was the "real starting point of the ENIAC project", however, he also states that "In fact, the ENIAC project was started only in May 1943, …" and Brainerd says: "Although it was dated August 1942, no document bearing that date was ever found". Arthur W. and Alice R. Burks report that "Work started on May 31, 1943" and Brainerd says: "work began officially in the Moore School on June 1, 1943". According to the Burkses "The ENIAC solved its first problem in December of 1945". This problem consisted of calculations for the development of the hydrogen bomb at Los Alamos. The first "working days" of ENIAC are described in more detail by Haigh, Priestley & Rope.[46] Colossus is usually not seen as a universal computer but rather as a special electronic computing device or as the "Godfather of the Computer" (Brian Randell), but there are also voices which claim that Colossus-1 was the first large-scale electronic digital computer. On the other hand, a trial to let Colossus perform base-10 multiplication was not successful. More details about the computational capabilities of Colossus are given by Haigh & Priestley in their paper on "Colossus and Programmability". The Burkses gave their book on Atanasoff the title "The First Electronic Computer", but they also clearly state that Atanasoff's device was a "special-purpose digital machine" and that ENIAC was " the world's first general-purpose electronic computer" and in 2018 ENIAC is called "one of the earliest electronic general purpose computers ever made."[47]

Especially the beginning of these unique projects cannot always be determined in a clear-cut way. There may be several specific possible dates, as e.g. 5 April 1954 vs. 17 May 1954 for the Oprema, or August 1942 vs 1 June 1943 for ENIAC. The memo of Howard H. Aiken, which led to the ASCC/Mark I, was apparently written in November 1937, more than one year before the approval by IBM's president.[48] Furthermore, preliminary work by and the at that point current knowledge of the people, which created these computers, may have had an influence on the time needed for development and construction. It has been pointed out above that Kämmerer obviously had done significant design work before 5 April 1954, and during his presentation Kortum said to minister Rau that the delivery of the necessary relays had already been secured.

It cannot be determined what specific knowledge about computers Kämmerer had, but we may assume that he had at least read the booklet of Heinz Rutishauser et al., which is mentioned in LWD on 24 April 1954: "Heft über „Programmgesteuerte digitale Regengeräte" erhalten [Have obtained booklet on „Program-controlled digital computing devices"; instead of „Regengeräte" it should read „Rechengeräte" because „Regengeräte" = „rain devices" obviously makes no sense]". This publication by Rutishauser, Ambros Speiser and Eduard Stiefel, who had visited the centers of computer development in the USA and the UK and also had close relations with Zuse, was praised by Franz L. Alt in 1952: "It is the first publication of its kind in German and far better than most similar writings in English. . . ., it is highly recommendable as an introduction to the subject." It contains a bibliography of about 70 items, six of which were not publicly available.[49]

Brainerd said in 1976: "When one looks into the background of the ENIAC, one finds at least ten different developments which prepared the Moore School to undertake the first large-scale electronic digital general-purpose computer;". He mentions e.g. earlier work of Chedaker, Eckert, and Sharpless "on advanced electronic circuits, electronic counters, and delay lines on project PL;" and "Burks and John Mauchly were involved in related radar work on project PZ (a Signal Corps project obtained by Brainerd);". Thomas P. Hughes reports that Mauchly "had with him a small breadboard model of a computer using neon tubes", when he visited and then joined the Moore School in 1941. In the light of the detailed analysis of the Burkses it is questionable whether this is an adequate description of the artifacts Mauchly had built at Ursinus College, and these artifacts are also not mentioned by Brainerd in 1976.[50]

Thomas H. Flowers, who headed the development and construction of the Colossi at Dollis Hill, says of himself: "By 1939 I felt able to prove what up to then I could only suspect: that an electronic equivalent could be made of any electromechanical switching or data-processing machine."[51]

Furthermore, these computers differed considerably in scale. ENIAC contained 18,800 vacuum tubes, Colossus-1 contained 1,500 tubes and Oprema-1 contained 8,313 relays.[52]

Last but not least, the circumstances under which these projects took place may have influenced the time for their realization. The Colossi were built in a country which was directly attacked by an enemy: "DH had its share of air-raid warnings. On one occasion DH was narrowly missed by a V2 rocket." DH stands for Dollis Hill, the British Post Office Research Station in London, were the Colossi were developed and built. Such circumstances implied enormous pressure on the people building Colossus-1. Harry Fensom, one of them, recalls: "It took us many months to build Colossus, working continuously, day and night. We even had a resident hairdresser to save valuable time!".[53] K&K and the Oprema team were under pressure due to the promise Kortum had given to Rau, who was a minister in a dictatorial state. The main reason for the ENIAC project was the necessity to compute firing tables for newly developed military weapons. The Moore School had recruited over 100 women with college degrees who computed such tables by means of paper and pencil and desk calculators, "yet, they continued to fall behind the work load".[54] Only the ASCC/Mark I project seems to have not been under such specific kinds of pressure. Aiken had been called to active duty in April 1941 and then worked as an instructor at the Naval Mine Warfare School in Yorktown, Va. At IBM "The war slowed progress on the Harvard machine, …" and "During World War II — while IBM factories were producing carbines, automatic rifles, cannon, bomb sights, gun directors, and fire control mechanisms, as well as punched-card machines of prewar design — IBM engineers undertook about a hundred special development projects for the armed forces."[55]

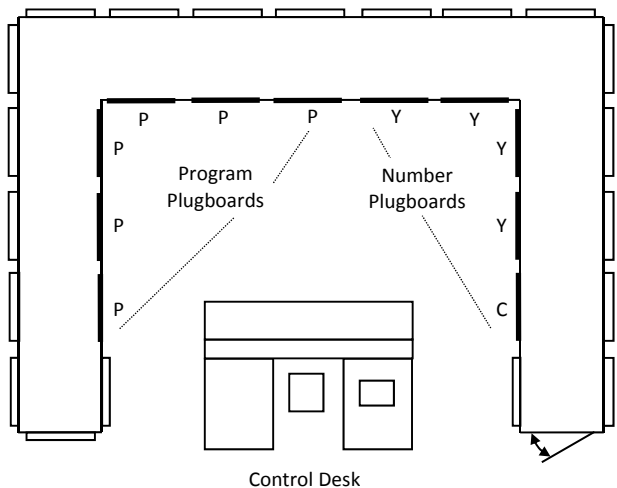

Figure 3. Floor Plan of One Oprema Machine (drawn by the author).

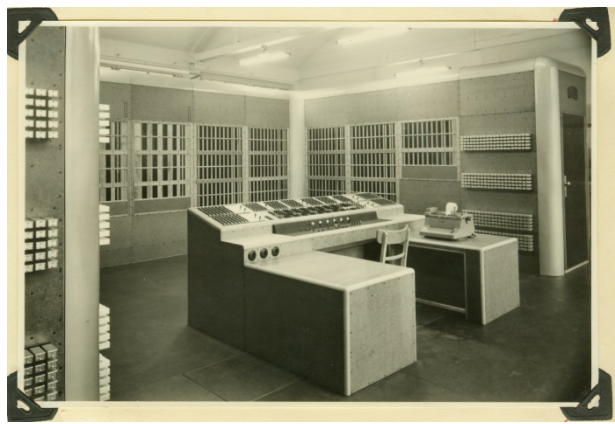

Figure 4. One of the Oprema Machines.
Source: ZEISS Archiv.

## Technical Properties of the Oprema

### Available Material

The Oprema was dismantled in the autumn of 1963 and the components were discarded or given away to people interested in them. It seems that the technical documentation suffered a similar fate. Apart from the work reports by Lenski there only exist some technical notes about general principles, the pulse generator and the arithmetic instructions, a paper by K&K, three papers by Kämmerer, 10 pages in the first edition of Kämmerer's book on "Ziffernrechenautomaten [Digital Computing Automata]" and a short operating instruction.[56] This is far less material than the "hundreds of pages" mentioned by Alt for the Bell Relay Computer Model V, or the material contained in the patent application Z391 of Zuse or that in the ENIAC patent. Thomas Haigh, Mark Priestley and Crispin Rope say that they "combed through mountains of archival materials" when preparing their book " ENIAC in Action".[57]

The Oprema was a relay computer were electromechanical relays were used for essentially all switching and for dynamic storage. One machine contained 8,313 relays, approx. 45,000 selenium rectifiers, approx. 250 km wire, about 500,000 solder joints and an unknown number of resistors.

### Physical Structure

Similar to the ENIAC *one* Oprema machine was a U-shaped assemblage of racks with the control desk inside the U. Fig. 4 shows one of the machines, and Fig 3 is a floor plan (drawn by the author) indicating the different groups of plugboards. The perimeter of one machine was 36.5 m. The panels on the inside contained the plugboards and some relays, and on the panels on the outer side the bulk of the relays was mounted in groups of 100 each. As can be seen in Fig. 4 all relays had a dust cover, which avoided malfunction due to dust and dirt. No incident with a moth similar to the famous "first bug", which Grace Hopper found in the Mark II relay computer on 9 Sep. 1945, was reported for the Oprema.[58] On the left arm there were 6 program plugboards (P) and on the right arm 5 plugboards for input numbers (Y and C). The role of these number plugboards will be detailed below.

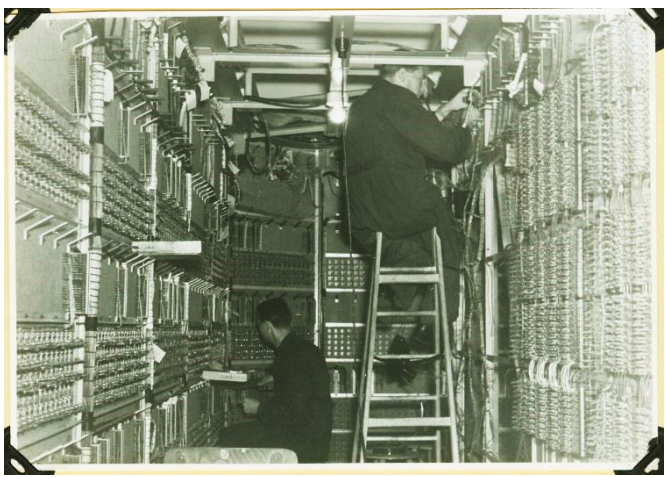

Figure 5. People working inside Oprema on 15 Dec. 1954.
Source: ZEISS Archiv.

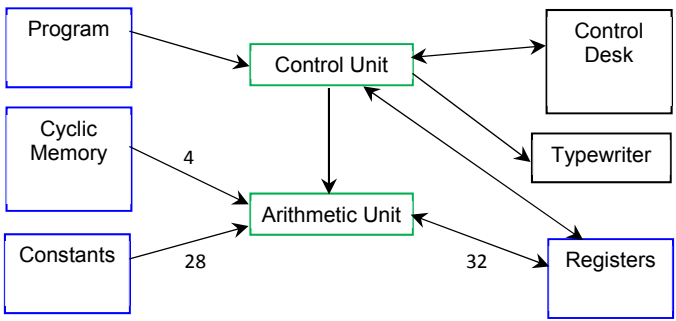

Figure 6. Block Diagram of the Oprema. (drawn by the author)

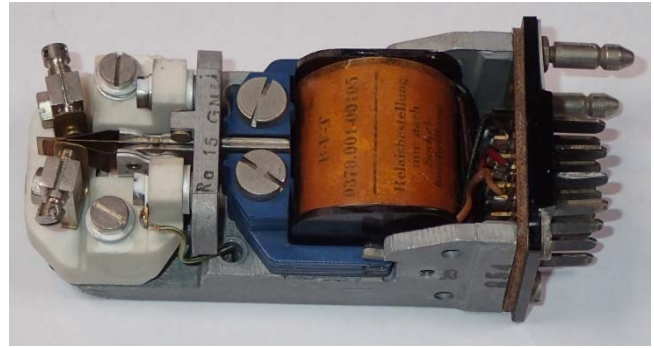

Figure 7. A relais of the Oprema. Photo: J. Winkler.

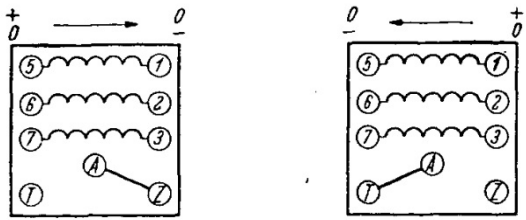

a: Spulenanschlüsse, Stromrichtung und resultierende Zungenstellung

Figure 8. Relay states and direction of current.
Part of Fig. 1 in Kämmerer's "The Relay Technology ...".

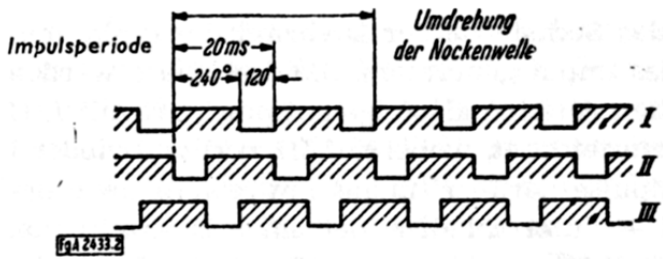

Figure 9. The 3−Phase Pulse System of the Oprema. Fig. 2 of Kämmerer's "The Relay Technology ...".

The wiring was in the interior of this U-shaped structure. Figure 5 shows two people working on the wiring in the interior room of one machine.[59] The corridor inside the machine was 1.1 m wide and 2.4 m high. The framework of the Oprema consisted of steel and the panels of lacquered high-density fiberboard. Lenski reports on 22 May 1954: "Anstatt Duralblech kann Holzfaserhartpappe genommen werden wenn es geht etwas stärker als 4 mm. [Instead of duralumin sheet we may use high-density fiberboard if possible a bit thicker than 4 mm.]"

### Logical Structure/Block Diagram

The logical structure of the Oprema is depicted in Fig. 6. The memory almost entirely consisted of plugboards, which were a kind of read-only memory. There were six plugboards for the program (P), each consisting of 50 rows, where one row contained one instruction. This means, that a program could consist of 300 instructions at most, which were numbered 0–299. Input data was contained in the plugboards for constants (C) and in the so-called cyclic memory (Y). The plugboard for constants consisted of 28 rows, where each row could store one number à 39 bits. The cyclic memory consisted of eight plugboards of 40 rows each, where each row consisted of 41 sockets, 39 for the 39 bits of a number and 2 sockets for the realization of shorter cycles. One cyclic memory consisted of two of these plugboards and therefore had a capacity of 80 numbers. The 32 internal registers consisted of relays and could store one number à 39 bits each. These registers were the only read-write memory of the Oprema.

It is interesting to determine the storage capacity of one Oprema machine:

| | | | |
|---|---|---|---|
| program: | $300 \times 27$ | = | 8,100 bits |
| cyclic memory: | $320 \times 41$ | = | 13,120 bits |
| constants: | $28 \times 39$ | = | 1,092 bits |
| registers: | $32 \times 39$ | = | 1,248 bits |
| | | | 23,560 bits |

Given these data, one Oprema machine had a storage capacity of approx. 3 kByte, excluding the temporary registers of the control unit and the arithmetic unit.

### The Relays

The relays were bistable polarized relays (latching relays) with up to three coils, each of which consisted of 5400 turns and had a resistance of 1200 Ω. Fig. 7 shows a relay which has come into the author's possession. Depending on the direction of the current through a coil the relay either assumes the state AZ or AT (see Fig. 8). If the current is turned off the relay remains in its present state. The relays were operated by direct voltage of ± 6 V, and a pulse of 5 ms duration was sufficient to put the relay in the corresponding state. In the following, a relay will be called "activated" if its contact is in position AZ and "deactivated" if its contact is in position AT.

## Overall Control

The machine was controlled by a system of pulses. The main circuits were controlled by a 3-pulse system (I, II, III), and the part that controlled the electric typewriter was controlled by a 2-pulse system. The 3-pulse system is depicted in Fig. 9. One pulse is $13^1/_3$ ms long followed by a pause of $6^2/_3$ ms, and the three pulse sequences are staggered symmetrically. The pulses are produced by cam-operated contactors in the pulse generator. For the control of relay systems with different pulse systems CZJ obtained a patent in the FRG (West Germany) but apparently not in the GDR (East Germany). The inventors named in the patent were Kämmerer, Kortum and Pöll.[60]

One design rule was that the coils of a relay were all driven by the same pulse sequence, i.e. with respect to its coils, a relay belonged to one of the pulse sequences I, II or III. A second design rule was that the contact of a relay whose coils were driven by pulse sequence I controlled the coils of relays of group II; the contact of a relay of group II controlled the coils of relays of group III, and the contact of a relay of group III controlled the coils of relays of group I. There were thus three groups of relays 1, 2, and 3 whose coils and contacts were associated with the three pulse sequences as indicated in table 1. An example of these design rules is the so-called "Führungskette [guiding chain]") which is detailed below. In the following I will use the term "control sequence" for Kämmerer's German term "Führungskette".

|  | Associated Pulses | |
| --- | --- | --- |
| Relay group | Coils | Contacts |
| 1 | I | II |
| 2 | II | III |
| 3 | III | I |

Table 1. Association between relay coils, relay contacts, and the three pulse sequences.

A consequence of these two design rules and the structure of the 3-pulse system is that the relay contacts are switched in a current-free state. This is an important factor for the reliability of the computer because it extends the lifetime of the contacts and also significantly increases their reliability. This had already been observed by Zuse around 1940. When a contact is opened while a current is flowing through it and thus a circuit containing any coils is broken a spark is produced by the self-induction of the coils. Such sparks lead to an increased contact wear-out which is especially important for the relays in computer circuits where the relays are switched much more often than in telegraphy and telephone systems. This problem had already been pointed out by Leonardo Torres y Quevedo in 1915.[61] The Oprema people did several endurance tests for the switching of the relays with contacts made of a gold-nickel alloy. The results were that switching under a load of

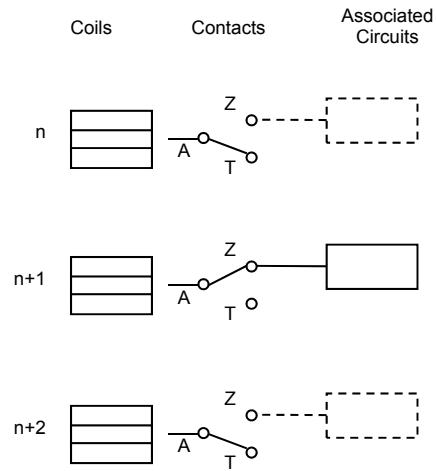

Figure 10. Basic principle of the control sequences. The activation wave has reached relay n+1 (drawn by the author).

200 mA led to wear-out of the contacts after 25 to 100 million switches, whereas the contacts were still in very good shape after one billion switches when switched in current-free mode.[62]

Fig. 11 shows the overall logical structure of the Oprema, and especially the different control sequences which controlled the sequencing of the single steps of the computation. Whereas Zuse used stepping switches for this control, the Oprema used relays for this purpose, too.

## Control Sequences

The control sequences are chains of relays which could split into branches and also merge again. In all paths the relays of the control sequences are numbered according to their distance to relay 1. The relays of a control sequence are activated one after the other, and, as a consequence, a wave of relay activations is flowing through the network. As can be seen in Fig. 9, a new pulse starts every $6^2/_3$ ms, which means that the activation wave is flowing through the network with a frequency of 150 relays/s. When a relay is activated it may trigger some circuit of the computation proper. This basic principle is depicted in Fig. 10. The activation wave has reached relay n+1 which triggers its associated circuit. Relay n is deactivated again and relay n+2 is still deactivated, and therefore they both do not trigger their associated circuits.

Due to the switching characteristics of the bistable relays and the control by the 3-pulse system Kämmerer used a somewhat more complicated scheme for the circuits of the control sequences. In this scheme, when relay n is activated, it

a) activates relay n+1,
b) triggers its associated circuit and
c) deactivates relay n−2.

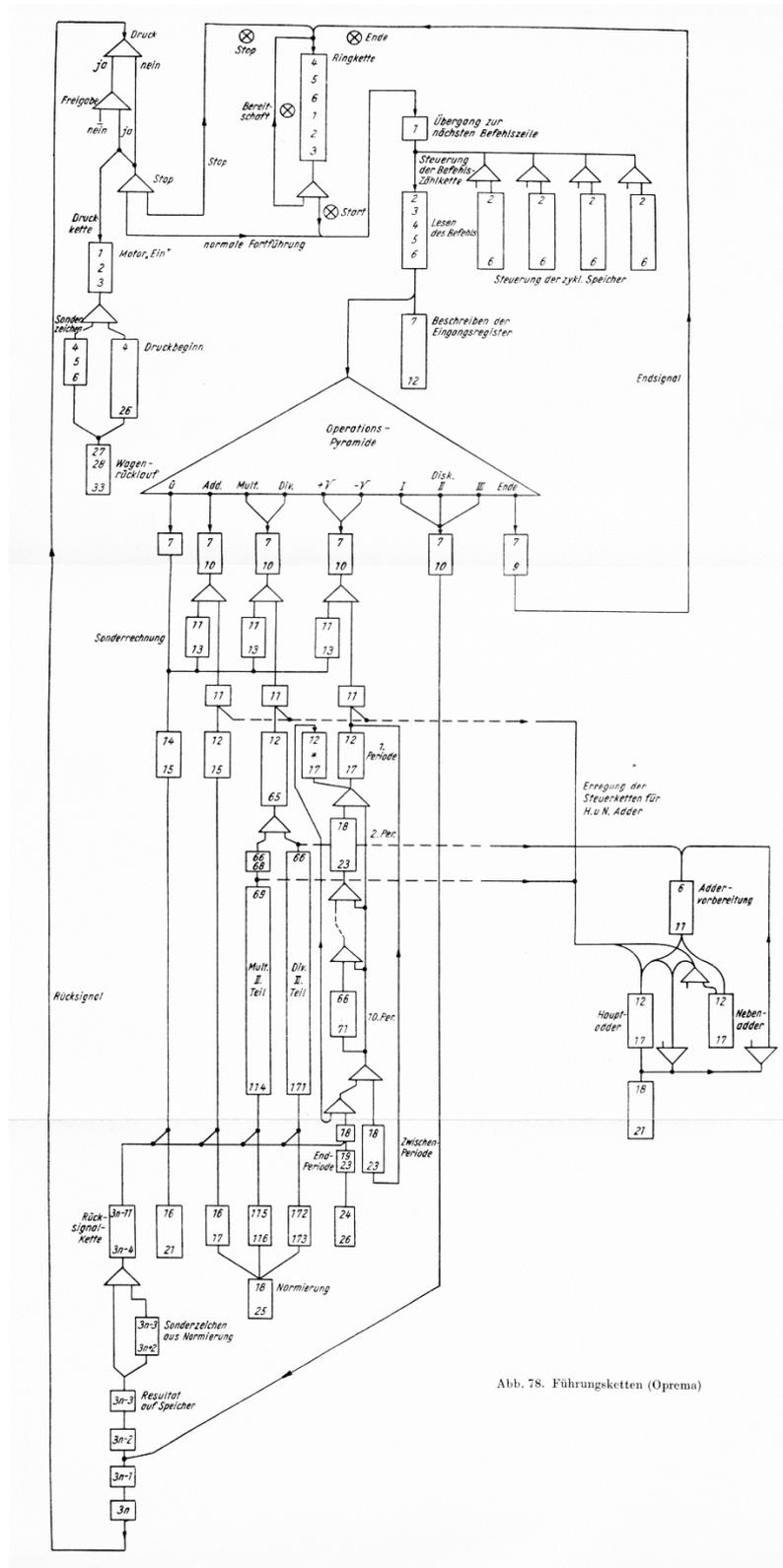

Abb. 78. Führungsketten (Oprema)

**Figure 11.  The Control Sequences of the Oprema. Fig. 78 from Kämmerer's "Ziffernrechenautomaten".**

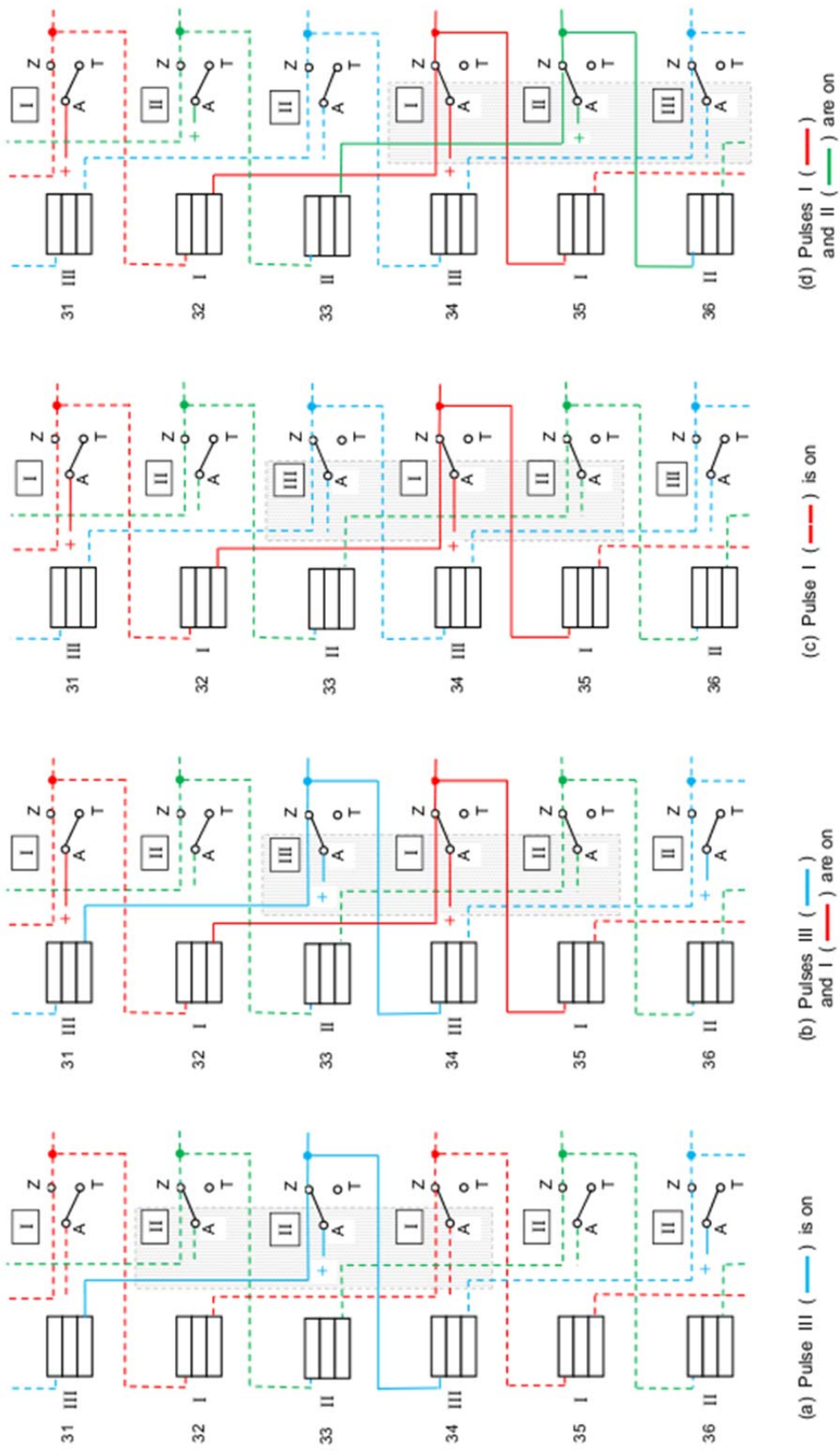

Figure 12. The sequential mode of operation of the control sequences (drawn by the author).

This is depicted in Fig. 12, in which the circuits associated to the relays of the control sequence are not shown. (Fig. 10 and Fig. 12 have been drawn by the author on the basis of Kämmerer's description in his 1958 paper on the relay techniques used in the Oprema.) In Fig. 12 solid lines carry a positive voltage and dashed lines are on ground level. The different colors indicate the three different pulse sequences: red for pulse sequence I, green for II and blue for III. The positive voltage of the pulses is fed into contact A of the relay. In this scheme the activation wave consists of three consecutive relays which are activated (AZ closed), while the other relays of the control sequence are deactivated (AT closed).

Fig. 12 shows six consecutive relays of a control sequence in four consecutive states. In order to understand these four states, we have to take into account a further detail of the 3-pulse system which is not explicitly depicted in Fig. 9: when the pulses are switched on or off there are very short time intervals during which only that pulse is on which does not change. E.g. if pulse I goes up and pulse II goes down only pulse III is on. This is depicted in Fig. 12 in state (a).

In state (a) relays 32, 33, and 34 are activated, and, as already mentioned, only pulse III is up. When pulse I goes up, relay 34, which is already activated, activates relay 35, which switches from AT to AZ. Additionally, relay 34 deactivates relay 32, which switches from AZ to AT. As a result, the relays 33, 34, and 35 are now on, and the other relays are off. Thus, the group of three activated relays has moved one step ahead.

The transition from state (a) to state (b) shows that the relay contacts are always switched in a current-free mode: the contacts of both the relays 32 and 35 are connected to the pulse sequence II which is down during this transition.

When pulse III goes down the network enters state (c) in which the state of the relays is the same as in state (b), whereas contact A of relays 33 and 36 is on ground level.

When pulse II goes up the network enters state (d), in which the relays 34, 35, and 36 are activated, and the other relays are deactivated. Again, the group of three activated relays has moved one step ahead, i.e. each new pulse moves the group of activated relays one step ahead.

### Execution of Instructions

The execution of an instruction starts with the activation of relay 1 in the middle of the upper part in Fig. 11. Depending on the instruction one of the paths through the network is taken, and when the instruction is completed control returns via the relays 3n-1 and 3n at the bottom to the circuit which decides whether the result of the instruction should be printed. If the stop-button has been pressed control enters the control sequence in the middle of the upper part. This control sequence consists of six relays which are wired in a circular fashion. This control sequence implements the idle state of the Oprema, in which the activation wave circulates in this control sequence. When the start-button is pressed relay 1 is activated and the execution of the next instruction begins. If, after completion of an instruction, the stop-button is not pressed control directly returns to relay 1.

When the machine should be switched off completely, the camshaft, which generates the pulses of the pulse systems, is stopped first and then the power voltage is switched off.

### Number Representation and Arithmetic

Numbers are represented as normalized floating-point numbers $\pm s \times 10^{\pm e}$, where

  s = Ddddddddd
  D is a decimal digit greater zero
  d is a decimal digit
  e is an integer with $0 \le e \le 15$.

The excess-3 code is used for the representation of the digits of the significand, i.e. each digit is represented by a 4-bit group, which Kämmerer called a "Tetrade [tetrad]". The exponent e is represented directly as a binary number using another tetrad. Together with the two bits for the two signs one number needed 38 bits for its representation, which were numbered 1 to 38. Since zero cannot be represented as a normalized floating-point number there was an additional 39th bit which was used in the following manner: for the normalized numbers bit 39 is off ("0"), and if bit 39 is on ("L") the bits 36, 37, and 38 represent one of three special values ("Sonderzeichen"):

|              | 36 | 37 | 38 | 39 |
|--------------|----|----|----|----|
| zero         | 0  | 0  | L  | L  |
| infinite     | 0  | L  | 0  | L  |
| indeterminate| L  | 0  | 0  | L  |

Kämmerer says that these special values are also used in subsequent computations, but the available documentation does not contain the algebraic rules for these values. These rules should be the same as those reported by Stiefel for Zuse's Z4:

  e.g.    infinite – infinite = indeterminate.

Zuse, who also used normalized floating-point numbers in his early computers, used the special values 0, $\infty$, and ? (indeterminate). Such special values are also present in modern floating-point formats as e.g. $\pm \infty$ and NaNs in IEEE 754-2008. Thus, the floating-point formats of Zuse and Kämmerer were early forerunners of the modern floating-point standards.[63]

Plugboards were used for the input of numbers. These plugboards consisted of rows of 39 sockets according to the 39 bits used for the representation of one floating-point number. There existed special four-pin plugs ("Tetradenstecker") for the different decimal digits of the significand. The sockets were interconnected in such a way that no plug was necessary for the digit zero. The numbers could also be plugged in in unnormalized form and would afterwards be automatically normalized when read into internal registers. There were two kinds of number plugboards whose role will be discussed in the section on programming.

The arithmetic operations of the Oprema were addition, subtraction, multiplication, division and extracting the square root. Addition was performed in a parallel manner because this was faster than serial addition. This compensated a little bit the disadvantage of using the rather slow relay technology. The arithmetic operations were all based on addition, and therefore the two parallel adders were the central components of the arithmetic unit. These two adders processed the significands of the operands. In the following it is assumed that the registers of the adders were 9 digits long, because this is necessary to add all significands correctly: e.g.

$$5000\ 0000 + 6000\ 0000\ =\ 1\ 1000\ 0000$$

But the available documentation does not describe such details of the adders. In the following examples the significands will be in this 9-digit format:

$$d_0\ d_1 d_2 d_3 d_4\ d_5 d_6 d_7 d_8\ .$$

For normalized significands

$$d_0 = 0\ \land\ d_1 > 0$$

holds.

Multiplication was performed in two phases. In phase 1 the nine multiples $s \times 1$, $s \times 2$, …, $s \times 9$ of the significand $s$ of the multiplier are computed by repeated addition and stored into the nine registers of the "Vielfachenspeicher" (storage of multiples). In phase 2 the significand of the product is computed by repeated addition of the multiples of the significand of the multiplier according to the digits of the significand of the multiplicand. This sequence of additions starts with the rightmost digit $d_8$ of the significand of the multiplicand. After each addition the partial product is shifted to the right by one place. Thus, the nine most significant digits of the significand of the product are computed. Parallel to this computation the two exponents are added together in the separate exponent adder and the sign of the result is determined by the signs of the two significands. Finally, the resulting sign, significand and exponent form the result, which is, in a last step, normalized into the form described above. This description of the multiplication follows that in the anonymous and undated note "Der Multiplikationsablauf" (cf. endnote 56). In Kämmerer's "Ziffernrechenautomaten" of 1960 the roles of multiplicand and multiplier are interchanged (p. 126), but since multiplication is commutative both descriptions are logically equivalent. Both descriptions are incomplete in that they do no describe how the mechanism works for the digit $0$ of the significand of the multiplicand resp. multiplier. Richard K. Richards calls this form of multiplication "N-tupling".[64]

The division operation consisted also of two phases. Phase 1 was similar to the first phase of multiplication: the nine multiples $s \times 1$, $s \times 2$, …, $s \times 9$ of the significand $s$ of the divisor are computed by repeated addition and stored into the nine registers of the storage of multiples. In phase 2 the digits of the significand of the quotient are computed sequentially from left to right by comparing the current rest $r$ – which is initially the significand of the dividend – with the multiples of the significand of the divisor computed in phase 1. The largest multiple $s \times i$ which is less or equal to $r$ is the next digit of the quotient:

$$s \times i\ \leq\ r\ <\ s \times (i+1) \tag{1}$$

This is basically the same algorithm as that of the usual pencil-and-paper method. This basic algorithm was improved in two points: (a) the comparison to determine the largest multiple which is less or equal to the rest was limited to the first two digits $d_0 d_1$ of both the rest and the multiples and (b) instead of the maximal $i$ with

$$(s \times i)_{0-1}\ \leq\ r_{0-1}$$

the minimal $i$ for which

$$(s \times i)_{0-1}\ \geq\ r_{0-1} \tag{2}$$

holds was computed.

Condition (2) is simpler to implement but has the drawback that $i$ is not in all cases the next digit $q_i$ of the quotient. If

$$(s \times i)_{0-1}\ >\ r_{0-1}$$

holds then

$$s \times i\ >\ r$$

does also hold and $q_i = i - 1$ is then the next digit of the quotient. But even if

$$(s \times i)_{0-1}\ =\ r_{0-1}$$

holds

$$s \times i\ >\ r$$

may also hold if

$$(s \times i)_{2-8}\ >\ r_{2-8}$$

is true.

This can be seen in the following example: let $s = 0\ 2010\ 0000$, $r = 0\ 8000\ 0000$, $i = 4$. With this we have: $s \times i = 0\ 8040\ 0000$ and

$$(s \times i)_{0-1}\ =\ r_{0-1}\ \equiv\ 08 = 08$$

but also

$$(s \times i)_{2-8}\ >\ r_{2-8}\ \equiv\ 040\ 0000 > 000\ 0000$$

and therefore

$$s \times i\ >\ r\ \equiv\ 0\ 8040\ 0000\ >\ 0\ 8000\ 0000\ .$$

In this case $i - 1$ is the correct value of the next digit of the quotient.

Therefore, after $i$ is determined according to (2) the two subtractions

$$ma := r - s \times i;\quad \text{and}\quad sa := r - s \times (i-1);$$

are performed in parallel in the main adder (ma) and in the secondary adder (sa), respectively. The new rest and the next figure $q_i$ of the quotient are then determined as follows:

> **if** $ma \geq 0$
> **then** $r := ma$; $q_i := i$;
> **else** $r := sa$; $q_i := i - 1$;
> **fi**

Similarly as in multiplication, the difference of the exponent of the dividend and the exponent of the divisor is computed in the exponent adder and the resulting sign of the quotient is determined by the signs of the two significands. In a last step the resulting sign, the significand of the quotient, which may be denormalized, and the resulting exponent form the result, which is normalized into the Oprema floating-point format.

Again, this presentation follows the description of the division operation in the anonymous and undated note "Der Divisionsablauf" (cf. endnote 56), but is a little bit more detailed than that. In his "Ziffernrechenautomaten" of 1960 Kämmerer gives a slightly more intricate description especially of the to the first two digits limited comparison of the current rest and the multiples of the significand of the divisor. Both descriptions are incomplete in that (a): $r$ must be multiplied by 10 after $q_j$ has been computed and (b): the registers in the adders and the storage of multiples should be 9 digits long.

It is interesting to note that Charles Babbage, who used fixed-point numbers, envisaged a very similar algorithm for division for his Analytical Engine. He also limited the comparison of the rest $r$ and the multiples of the divisor $d$ to the first two digits, respectively. As the candidate for the next digit of the quotient he chose the largest multiple $d \times i$ which is less or equal to $r$. With this we obtain the kernel of Babbage's algorithm:

> $i := \max(k: (d \times k)_{0-1} \leq r_{0-1})$;
> $r := r - d \times i$;  $q_j := i$;
> **if** $r < 0$
> **then** $r := r + d$;  $q_j := q_j - 1$;
> **fi**

It seems improbable that Kämmerer knew of Babbage's algorithm, and thus, this may be an example for the conclusion which Maurice Wilkes draws at the end of his paper "Babbage as a Computer Pioneer": "As it is, everything that he discovered had to be re-discovered later."[65]

The most complicated arithmetic operation of the Oprema was the extraction of the square root. It is described in the short, anonymous and undated note "Das Wurzelziehen" (cf. endnote 56) and in more detail in Kämmerer's "Ziffernrechenautomaten" of 1960 (pp. 110−111, 128−129). If the radicand is negative the result is the special value "indeterminate". For positive radicands the operation is based on the well-known relation

> $n^2 = 1 + 3 + \ldots + 2n{-}1$

i.e. $n^2$ is the sum of the first $n$ odd integers. A basic algorithm implementing this relation works by successively subtracting consecutive odd integers from the radicand resp. rest beginning with 1. If the radicand is somewhat larger this basic algorithm leads to a sizable number of subtractions. The computation of e.g. $\sqrt{1225} = 35$ involves 35 subtractions and 35 additions to compute the sequence of odd integers. In the Oprema the basic algorithm is improved by applying these subtractions to groups of two decimal places as it is done in the well-known paper-and-pencil "division" method. With this improvement the computation of $\sqrt{1225}$ is completed in 11 steps. This improved method is also described by Richards,

but it seems that he was not aware of the Oprema.[66]

### Rounding

Due to the fixed length of the significand in the Oprema floating-point format the significand of the result of an arithmetic operation had to be rounded to fit the Oprema format described above. The Oprema used two forms of rounding, one in the main adder and a second in the normalization group. In the main adder rounding is necessary after right shifts, when e.g. during addition the significand with the smaller exponent must first be shifted right n places, where n is the difference between the larger and the smaller exponent of the two summands. Right shifts also occur during multiplication, as already mentioned above. In these cases the rounding method is roundTiesToAway aka banker's rule which was the standard rounding in numerics at the time of the Oprema.[67]

In the normalization group a simpler rounding method was employed (roundNorm), which seems to be rather unique. The method takes the last two digits $d_8 d_9$ of the unrounded significand

> $s_u = d_0 d_1 d_2 d_3 d_4 d_5 d_6 d_7 d_8$

into account according to the following rule:

> $d_8 = 0$ $\mapsto s_r = 0 d_0 d_1 d_2 d_3 d_4 d_5 d_6 d_7$
> $d_8 \neq 0 \wedge d_7$ is odd $\mapsto s_r = 0 d_0 d_1 d_2 d_3 d_4 d_5 d_6 d_7$
> $d_8 \neq 0 \wedge d_7$ is even $\mapsto s_r = 0 d_0 d_1 \ldots d_6 (d_7{+}1)$

where $s_r$ is the rounded significand.

Examples:

> roundNorm($1\ 2222\ 2219$) $= 0\ 1222\ 2221$
> roundNorm($1\ 2222\ 2221$) $= 0\ 1222\ 2223$

The reason for this special rule was that $d_7{+}1$ will never produce an overflow because $d_7 = 8$ is the largest possible even value. In roundTiesToAway rounding is usually done by adding 5 to $d_8$ which may result in an overflow into $d_7$ and this overflow may ripple to the left until $d_0$ as in

> $s_u = 0\ 9999\ 9995$,

which means that for roundTiesToAway a full adder is necessary. RoundNorm can be implemented by a much simpler circuitry and this is mentioned as the reason in the anonymous and undated note "Abrundung bei der Normierung [Rounding during normalization]" (cf. endnote 56). In Kämmerer's "Ziffernrechenautomaten" of 1960 the description of this rounding method contains a contradiction and therefore seems to be incorrect.[68]

The simpler circuitry also has some disadvantage: the mean absolute rounding error is about twice that of roundTiesToAway, which is the same as that of the modern roundTiesToEven. On the other hand, the mean rounding error is zero, i.e. when summed up over the 100 possibilities of $d_7 d_8$ the rounding errors are cancelled out and, therefore, roundNorm shows no drift in the sense of John F. Reiser and Donald E. Knuth.[69] With re-

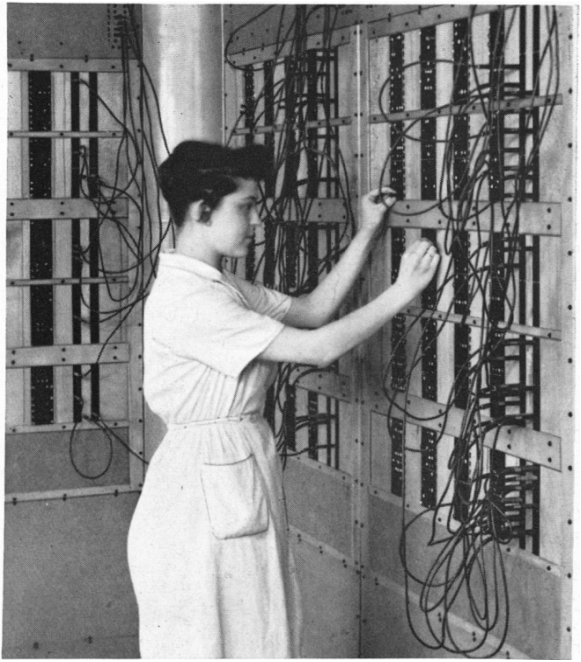

**Figure 13. "Loading" of a program at the plugboard. Source: de Beauclair "Rechnen ...", p. 103 (https://archive.org/details/rechnenmitmaschidebe).**

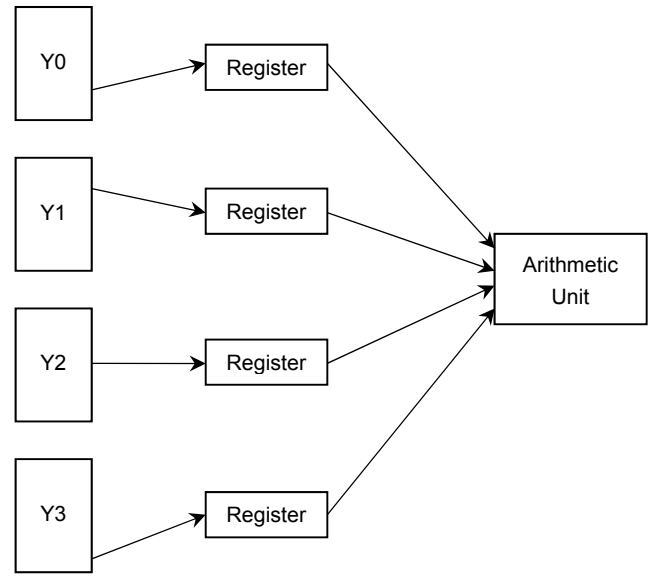

**Figure 14. The four cyclic memories (drawn by the author).**

spect to the ambivalent case $d_8 = 5$ roundNorm behaves as roundTiesToOdd. RoundTiesToEven (RTTE), which was recommended by James B. Scarborough in 1930 and which is attributed to Carl Friedrich Gauss by Burks, Goldstine and von Neumann in 1946[70], can lead to the rippling overflow mentioned above:

RTTE(1 9999 999**5**) = 0 2000 0000

As already mentioned above, the Oprema also had a secondary adder, but the remaining documentation contains no information about the rounding method used there. Since the secondary adder essentially had to perform additions of the same kind as the main adder it may be assumed that roundTiesToAway was also used in the secondary adder. Kämmerer remarks that the primary adder and the secondary adder had the same structure.[71]

**Instruction Set**

The Oprema was a four address machine with the following instruction format:

| Adr1 | Op | Adr2 | Adr3 | Adr4 |
|------|----|------|------|------|
| 6 | 6 | 6 | 5 | 4 | bits |

One instruction consists of 27 bits. Adr1 and Adr2 refer to the input data of the instruction and Adr3 refers to the place where the result is stored, where the following condition must hold:

Adr3 ≠ Adr1 ∧ Adr3 ≠ Adr2 .

The input can come from the constant plugboards (28 numbers), from one of the 4 cyclic memories or from one of the 32 internal registers. Together, this are 64 different possibilities, and therefore,

Adr1 and Adr2 are 6 bit wide. The result can be written to one of the 32 internal registers, and therefore, 5 bits are sufficient for Adr3. Adr4 consists of 2 × 2 bits and determines the next instruction to be executed. If these 4 sockets are empty the next instruction in the program plugboard is executed, where instruction 299 is followed by instruction 0. The first two sockets of Adr4 are used for conditional jumps and the last two sockets are used for unconditional jumps.

Adr4$_C$     Adr4$_U$

○   ○    ○   ○

cond.      uncond.      jump

For an unconditional jump from instruction i1 to the instruction i2 the left socket of instruction i1 and the right socket of instruction i2−1 are connected to each other by a jumper cable:

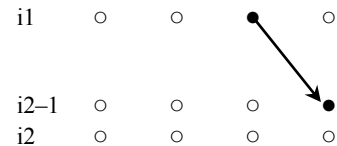

This holds for forward jumps as well as for backward jumps within one program table. For the realization of jumps between different program tables there exist special connections between the tables ("Fernverkehrsstraßen" ["long-distance highways"]). The destination of a conditional jump is determined in an analogous manner using the first two sockets of Adr4. Figure 13 shows the "loading" of a program at the plugboard.[72] Jumper

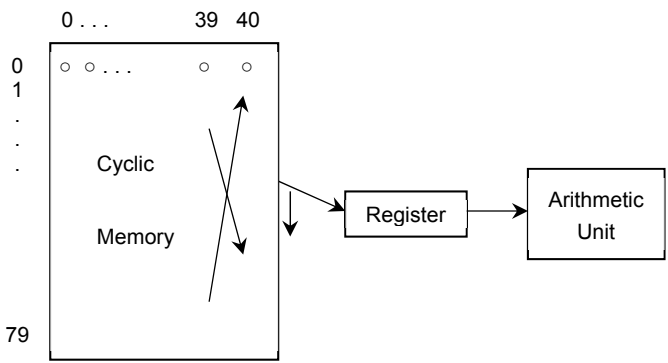

**Figure 15. A cyclic memory with two jumps (drawn by the author).**

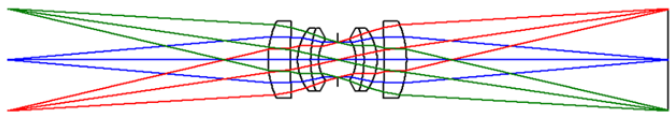

**Fig. 16. Ideal 1:1 Projection by a Symmetrical Lens System. Courtesy of Eberhard Dietzsch.**

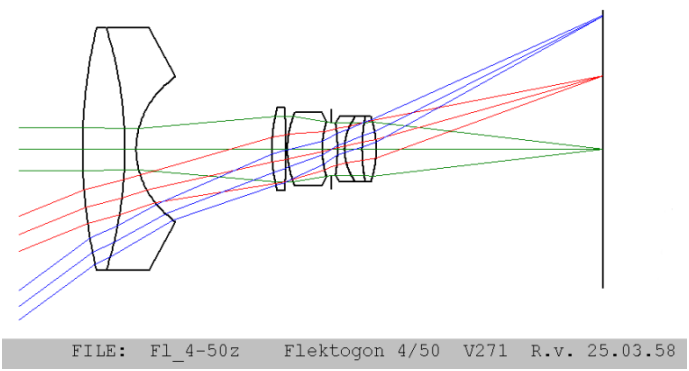

```
FILE: Fl_4-50z    Flektogon 4/50  V271  R.v. 25.03.58
```

**Fig. 17. The Retrofocus Wide-Angle Lens Flektogon 4/50 with Selected Ray Paths for an Infinite Object Distance. Courtesy of Eberhard Dietzsch.**

cables are seen in the two narrow columns at the right end of each program table.

The operation field is 6 bit wide, which allows for 64 different operations. The first five bits determine the operation proper and bit 6 controls the printing of the result of the current instruction. The remaining five bits allow for 32 operations, 25 of which were actually used.

There were 16 different kinds of addition, which varied in the treatment of the two operands (Adr1) and (Adr2) where, as usual, "(Adr)" means the number contained in the memory cell with address "Adr". Each of the two operands could come in four forms:

$$+ (Adr), \quad - (Adr), \quad + |(Adr)|, \quad - |(Adr)|,$$

and therefore

Adr3 := + |(Adr1)| + − |(Adr2)|;
print( (Adr3) ); goto 253;

could be expressed in one instruction. A very similar scheme was used in the ASCC/Mark I.[73]

The remaining 9 operations were:

Adr3 := (Adr1) × (Adr2);
Adr3 := (Adr1) / (Adr2);
Adr3 := + √(Adr2);
Adr3 := − √(Adr2);
**if** (Adr1) > 0 **then** goto Adr4$_C$;
**if** (Adr1) = 0 **then** goto Adr4$_C$;
**if** (Adr1) = ∞ **then** goto Adr4$_C$;
Adr3 := (Adr1);
end of computation (loop stop).

The Oprema was, as this instruction list shows, well suited for all kinds of numerical computations within the limits of its restricted storage capacity. The only feature which was specific for optical calculations was the so-called cyclic memory, whose logical structure is depicted in Fig. 14. As already mentioned, this consisted of four groups of two plugboards, where each of these plugboards contained 40 rows à 41 sockets, the first 39 sockets for the 39 bits of a floating point number and the last two sockets for the realization of jumps analogous to the unconditional jumps in the instruction tables. Access to the cyclic memories is provided by four registers which contain one of the numbers of the corresponding cyclic memory. The settings at the control desk determine which numbers are in these registers when the program execution is started. After each read access to one of the registers the next number of the corresponding cyclic memory is transferred into this register. This access is done in a cyclic manner. Additionally, it is possible to alter this basic cyclic order by "unconditional jumps" using the sockets 39 and 40 in the same way as the sockets 25 and 26 are used in the program tables (see Fig. 15).

The mechanism of the cyclic memories can be seen as a simple form of index registers. It was motivated by the calculations for the design of compound lenses, as e.g. camera lenses.

A camera lens should map the object, ideally, as an exact, i.e. geometrically similar, image onto the film or, nowadays, onto the image sensor. In order to do this it is necessary that all different light rays travelling from one point of the object plane through the several lenses of the lens system meet again at one corresponding point of the image plane. Fig. 16 shows the theoretical principle and Fig. 17 is a real design drawing by Eberhard Dietzsch, a former lens designer at CZJ. It is based on data of the design of the f = 50 mm Flektogon f/4 in 1958.[74] Fig. 17 is a computer generated diagram. Back in the 1950ies such diagrams were drawn by hand.

In reality, the ideal mapping can only be approximated, because, e.g. the refractive index of glass varies with the wavelength/color (chromatic aberration). If we want a picture which is sharp for objects in different planes things are even more complicate and rather impossible by purely optical means. One current approach to overcome these limitations is focus stacking.[75] At the time of the Oprema, geometrical ray tracing was often used in lens design. This means, that the paths through the lens system of e.g. 100 different rays are computed using the laws of geometrical optics. For each ray the refraction at the surfaces between two optical media, e.g. glass and air, or two different sorts of glass, has to be computed sequentially from object to image. The refraction at one spherical surface depends on the refractive indices $n_{1,2}$ of the two media, the radius $r$ of the surface and of the angle

and the distances were stored analogously:

| Y2[0] | Y2[1] | Y2[2] . . . |
| $d_0$ | $d_1$ | $d_2$ . . . |

These sequences were realized as circular sequences by means of the "jump" mechanism described above.

The characteristic values ($cv$) of the different rays, e.g. starting point and initial direction, were stored in the last cyclic memory:

| Y3[0] . . . | Y3[k] . . . | Y3[2k] . . . |
| $cv_0$ | $cv_1$ | $cv_2$ . . . |

were each $cv$ consisted of k numbers.

The number of rays and the number of surfac-of incidence $i$. Furthermore, the distance $d$ between two successive surfaces has to be taken into account. When all the paths of all rays have been computed, it is checked whether the quality of the image is sufficient. In the beginning of the design of a lens system this is often not the case. Then one or more of the parameters are changed by a small amount and the computation is repeated. This approximative process is continued until an acceptable design is obtained which is then used for production.

|  | Oprema | 1ˢᵗ computer | 2ⁿᵈ computer |
|---|---|---|---|
| a) Time for preparation | Programming 10 "Loading" of the program 20 | 100 | 100 min. |
| b) Time for calculations | 31 | 1100 Contains 80 min. for the recalculation of 15 erroneous results found by the computer himself. | 790 Contains 30 min. for the recalculation of 15 erroneous results found by the computer himself. |
| c) Number of errors found only in the final comparison | 0 | 19 on average almost 10 % unnoticed errors | 9 |
| d) Addition for debugging (200% per error) | 0 | 266 | 90 |
| e) Time for preparing and checking a typewritten transcript of the results | – | 90 | 90 |
| f) Factor of times for calculations and debugging (b+d) | 1 | 41 | 28 |

**Table 2. The results of Kämmerer's experiment.**

Kämmerer describes the use of the cyclic memories for these calculations in one example. The refractive indices were stored in the order from object to image in one of the cyclic memories, e.g. Y0:

| Y0[0] | Y0[1] | Y0[2] . . . |
| $n_0$ | $n_1$ | $n_2$ . . . |

the radii were stored in a second cyclic memory:

| Y1[0] | Y1[1] | Y1[2] . . . |
| $r_0$ | $r_1$ | $r_2$ . . . |

es were stored in the constant table, and thus, the machine was in a suitable state to begin the ray tracing calculations.[76]

### Reliability

One of the reasons K&K gave for the use of relays instead of vacuum tubes, which had then typically been used in computers for several years already, was that relays were more reliable than tubes.[77] But even the early electromechanical computers also had serious reliability problems:

"Computations were checked frequently: as often as every twenty minutes to ensure that the machine was not spewing out volumes of nonsense. Besides using that check, problems were often recorded and run again using the same direct evaluation method, but using different parts of the machine. In all, the checking of the machine for errors was an important, time consuming part of the work of the Havard Computation Laboratory."[78]

As has already been mentioned above, experiments had shown that the contacts of the relays were still in very good shape after one billion switches when switched in current-free mode. But nevertheless, the relays turned out to be the main source of problems during the debugging of the two machines in 1955. On one occasion almost all relays of Oprema-2 and the 2,000 relays in reserve had to be partially disassembled, readjusted and then reassembled, because a rivet of the armature had to be retightened.

Furthermore, Kämmerer did an experiment in 1955 during the trial period (May to July 1955) to assess the reliability of Oprema-1. It consisted of the computation of a polynome of $5^{th}$ degree for 151 argument values. These calculations were done by Oprema-1 on the one side and by two experienced (human) computers, working independently, on the other side. The human computers worked with paper, pencil and 8-digit Mercedes Euclid desk calculators. The results of this experiment are given in a table of Kämmerer (Table 2). In this table "computer" means a human computer, which was still a common term at that time.[79] The essential result was that the Oprema was *faster and better* than the human computers. The Oprema always computed the correct result. Kämmerer was especially astonished that the individual results of the human computers contained about 10% erroneous values despite the fact that the form of the polynome facilitated the checking of the values. Unfortunately, his report does not contain the formula of the polynome so that we cannot judge this aspect of the experiment ourselves.[80]

The original design of the Oprema envisaged a system of two identical machines which could work in two different modi: as two independent computers and as a combined system. In the combined modus the two machines would execute the same program with the same input data. After the completion of an instruction the two results would be compared. If they were different, the machines would enter the idle state mentioned above and a red and a green lamp at the console would be turned on. In this situation there would be several possibilities for continuation:

– repeat the last instruction; this could be successful in case of a transient failure;
– insert the correct result values and continue the program;
– execute some test programs in order to try to locate the fault, do the repair, and then continue the program;
– switch the machines off.

The twin-machine approach to improve the overall reliability of a computing system was also used in the Harvard Mark II and the BINAC. Both are mentioned in the booklet of Rutishauser et al., but no details are given there, and it is not known whether Kämmerer knew the more detailed publications which had been published at that time. The Mark II was even more versatile than the Oprema: additionally to the two modi described above, the Mark II could also work as one combined machine on problems too large for a single machine.[81]

Since both machines worked very reliably after the debugging was completed, the cables to connect the two machines and make the checking modus possible were never installed, and the Oprema was thus always used as two separate computers.[82]

In his report "5 Jahre "Oprema" [5 Years "Oprema"]" Alfred Jung gives the productive hours for both machines together for the years 1956 – 1960:

| 1956 | 8,589 | 64 % |
|---|---|---|
| 1957 | 9,445 | 70 % |
| 1958 | 10,512 | 78 % |
| 1959 | 11,530 | 85 % |
| 1960.Jan-Jun | 5,948 | 88 % |

The basis for these statistics are ca. 13,520 possible productive hours per year for both machines together: each week from Monday 06:00 to Saturday 16:00.[83]

## Operation and Use

On 1 August 1955 the "Rechenraum Oprema [Computing Room Oprema]" was founded as a separate unit, and Oprema-1 was put into productive use. Alfred Jung, a mathematician who had worked on the logic design of the Oprema, was the first supervisor of the Rechenraum. As for the date of the start of the project, different dates for the beginning of the productive use of Oprema-1 are mentioned by different authors. Kortum mentions 1 May 1955, Fritz Straube, a deputy of Kortum, mentions the beginning of June 1955, whereas Jung says in the report "5 Years "Oprema" " that the productive use started at 1 August, and Kämmerer says "von August an [from the beginning of August]".[84]

It seems that, despite the fact that the Oprema was retired in autumn 1963, the first five years were the most productive time of the Oprema, because two machines of the type ZRA 1 (Zeiss-Rechenautomat 1 [Zeiss Computing Automaton 1]) were installed in the Zeiss computing center in spring 1960, which boosted the capacity of the computing center by a factor of 40. The ZRA 1 was the second computer model developed and built by CZJ; about 33 ZRA 1 were produced. The "Computing Room Oprema" had been renamed "Computing Center" in 1959.[85]

It seems that after the Rechenraum Oprema had been founded there was still some debugging necessary. Kämmerer reports for Oprema-1 10 productive weeks from August to December 1955. During these weeks the machine was operated 132[86] hours per week: three shifts from Monday to Friday and two shifts on Saturday. During these 10 weeks 18 orders for optical systems had been completed in 1040 productive computing hours in total. In a separate table Kämmerer lists 21 orders with a total of 978 hours, which were billed to the customers with 115 DM per hour.[87] 15 of these orders came from WOPho, the computing bureau for photographic lenses, whose head was Harry Zöllner, who had also advocated the development of the Oprema. For WOPho the use of the Oprema was a great success. A table dated 1 January 1958 shows that the number of successive design samples during the development of one design ready for production dropped sharply after the Oprema had been used:[88]

| Year | 1954 | 1955 | 1956 | 1957 |
|---|---|---|---|---|
| a) # of design samples | 29 | 17 | 23 | 18 |
| b) # of designs ready for production | 2 | 7 | 16 | 15 |
| Percentage of b) | 7 | 41 | 70 | 83 |

Two reasons for these improvements were that more rays could be included in the calculation and that it was also possible to include aspheric surfaces.[89]

There also were a number of external customers, because during those years the Oprema was the sole program-controlled computer in the GDR. Therefore, the percentage of productive time used for orders from CZJ dropped steadily

| 1956 | 95 % | CZJ |
|---|---|---|
| 1957 | 90 % | CZJ |
| 1958 | 77 % | CZJ |
| 1959 | 75 % | CZJ |

Among the external customers were:
University of Rostock
VEB Turbines and Generators, Berlin
WTB Nuclear Reactor Construction, Berlin
German Federal Railway
German Metereological Service, Potsdam
Observatory Babelsberg
Geodetic Service, Leipzig
University of Halle.[90]

## Why relays in 1954?

Despite the fact that Kämmerer was well aware that vacuum tubes were already used in computers and that transistors were the switching components of the future, K&K decided to use comparatively slow relays as the switching elements of the Oprema. One reason was that relays were more reliable than vacuum tubes, a fact which is also mentioned by George R. Stibitz, who was heavily involved in the development of the relay calculators and computers at Bell Labs during the 1940s: "And so, with Model VI an era was ended. By 1950 the excitement of electronic potentialities was sweeping the country, and our blessedly reliable old relays were slow by comparison."[91]

With a multiplication time of 800 ms the Oprema was about as fast as other parallel relay computers, as e.g. the Bell Relay Calculator Model V (1 s) and Aiken's Mark II Calculator (800 ms). But the relay computers were significantly slower than their electronic counterparts, as e.g. ENIAC (5.6 ms). Konrad Zuse, who never saw the Oprema, supposed that it "was probably the fastest relay computer ever built."[92] But a closer look reveals that his own Z5 (completed in 1953) was slightly faster than the Oprema[93]:

| | Z5 | Oprema |
|---|---|---|
| Addition | 100 ms | 120 ms |
| Multiplication | 400 ms | 800 ms |
| Division | 750 ms | 800 ms |
| Square root | 750 ms | 1200 ms |

A second reason for the use of relays seems to be that K&K and CZJ as a whole were not familiar with electronics but rather with mechanical and electro-mechanical methods. The devices developed by Kortum and his group in the 1930s and 1940s consisted of mechanical, electro-mechanical and optical components. In their 1955 paper on the Oprema K&K mention that a computer for optical calculations based on analog elements had been designed at CZJ before. This machine would have had a number range too small for the intended applications, and they therefore decided to build a digital computer. Furthermore, the tight schedule based on Kortum's promise to minister Rau left no time to study and learn the new technology of digital electronics. Kortum still wrote in 1959: "Die bei uns vorhandenen Röhren sind solchen Anforderungen noch nicht genügend gewachsen, … [The vacuum tubes availabe for us do not yet meet such requirements, …]". Only in the 1960s was electronics introduced at Zeiss. The first internal conference on electronics took place in 1964, and when the last general director, Wolfgang Biermann, came to CZJ in October 1975, he observed that CZJ had begun to incorporate electronics into their products too late.[94]

The situation of K&K with respect to electronics was very similar to that of Aiken and his co-workers in 1944 and 1945 when the Mark II project was started. The Navy "urged that the machine be so designed that its construction could be completed as soon as possible. Accordingly, it was decided to build a relay calculator." Additionally, Robert Campbell, who was involved in the design of the Mark II relay calculator, later wrote in retrospect: "It may be noted that we had no one on board with experience in the technology of television or of radar: these fields were to provide the principal expertise in high-speed pulse circuitry." Eckert, one of the designers of ENIAC, is even more explicit in a letter to Robert P. Multhauf of the Smithonian Institution dated 22 July 1963: "I was chief engineer and thus in charge of the ENIAC project. My experience with analog computers was no great help in building the digital ENIAC, ... The influence of the radar switching and timing circuitry was more important and significant than the analog computer." Flowers, who headed the development and construction of the Colossi has already been cited above: "By 1939 I felt able to prove what up to then I could only suspect: that an electronic equivalent could be made of any electromechanical switching or data-processing machine."

On the other hand, both Heinz Billing and Maurice V. Wilkes (EDSAC) had experience in electronics when they began to develop electronic computers, and, had, by no means, such pressures as the Colossus, Mark II, ENIAC and Oprema people.[95]

## Overall evaluation

Despite the fact that technologically it was a late-comer, Oprema was the 7[th] computer in Germany and the first computer in the GDR.

The first computers in Germany were:

| | | |
|---|---|---|
| Z3 | Zuse | 1941.May |
| Z4 | Zuse | 1945.March |
| G 1 | Billing | 1952.Autumn[96] |
| Z5 | Zuse | 1953.July[97] |
| ALWAC | Logistics Res. | 1953.December (?) |
| G 2 | Billing | 1955.Spring[98] |
| Oprema | Kämmerer | 1955.Summer |

As has been discussed above in connection with the several dates regarding Oprema, ASCC/Mark I and ENIAC, it is not always easy to give precise dates for the different relevant events in the life of these early computers. For the Zuse computers I follow Zuse who said: "It [the Z3; JW] was completed in 1941 and was the first fully operational machine to contain state-of-the-art versions of all the important elements of a program-controlled computing machine for scientific purposes."; i.e. Zuse himself does not see his first two experimental models, Z1 (1938) and Z2 (1940), as computers. On the other hand, IFIP writes in 1994: "*Prof. Zuse,* who designed the Z1, "the first working programmable computer," in 1935-6,". Raul Rojas wrote that the Z1 "has been called the first computer in the world". The date for Zuse's Z4 is a presentation at the Aerodynamics Research Institute in Göttingen just before Easter 1945, when Zuse was on his way from Berlin to the Alps. In the following years it was occasionally used for different computations, e.g. in 1948 for commercial calculations for the alpine dairy cooperative Lehern in Hopferau, and its productive use began in August 1950 at the Institute for Applied Mathematics of the Swiss Federal Institute of Technology (ETH Zurich) as the first computer in Switzerland.[99]

The date for the ALWAC (Axel Lennart Wenner-Gren Automatic Computer) comes from a survey of the Office of Naval Research (ONR). The information in this survey "was obtained, for the most part, during the month of February 1953, …". This means, that "December 1953" seems to be an estimate made by Logistics Research, Inc. of Redondo Beach, California, early in 1953. On the other hand, the journal "Electrical Engineering" in May 1954 reports that the ALWAC has been unveiled, without giving any date. Logistics Research was formed in 1952 by the Swedish industrialist Axel Lennart Wenner-Gren, who had also founded the "Verkehrsbahn-Studiengesellschaft [Transit Railway Study Group]" in Cologne-Fühlingen, Germany, in 1951, which developed a monorail train system, the ALWEG-Bahn. In January 1953 the Verkehrsbahn-Studiengesellschaft was renamed "Alweg-Forschung, GmbH [Alweg Research Corpora-tion]". The ONR report mentions the "ALWEG Corp., Cologne, Germany" as one of the installation sites and so does the BRL report of 1955.[100]

It seems that Kämmerer made the best of the relay technology he and Kortum had chosen as the basic technology. They used bi-stable self-latching relays which were switched between two stable states by short pulses, and therefore did not need a permanent current to hold them in one or the other state. This reduced the power consumption significantly. Kämmerer reports several times that the power consumption of the Oprema was only 30 to 40 Watts for the machine alone, without the lamps at the control desk. He never mentions the electric motor which drove the cam shaft. It is therefore a bit difficult to compare this rather low power consumption with that of other relay computers; e.g. Z3: 4 kW, Z4: 4 kW and Z5: 5 kW. The motor, which drove the shaft of the ASCC/Mark I, consumed about 3 kW.[101] The relay contacts were always switched in current-free mode which reduced the wear of the contacts significantly; only the contacts of the cam shaft had to be checked regularly and maybe replaced.[102]

A peculiarity of the Oprema was the use of plugboards as program storage. It had the drawback that the "loading" of a program was rather cumbersome and error-prone. In his work diary Lenski reports that Kämmerer had the idea to employ plugging templates to facilitate the loading of programs, but it seems that such templates had never existed.[103] On the ENIAC, where programming may have been even rather more cumbersome, the Burkses remark: "While the ENIAC could compute the 30-second trajectory of a shell in 20 seconds, operators required 2 days to program it to do so."[104]

On the other hand, the program in the Oprema was thus stored in the machine, which allowed arbitrary conditional and unconditional jumps and therefore also the execution of loops.[105] As has been discussed above, the programs for optical calculations were of a highly repetitive nature. The down-time caused by program loading could be reduced by plugging-in a second program in the unused parts of the program tables while the machine was executing the first program.[106] This worked well as long as the total length of the several programs did not exceed the limit of 300 instructions.

## Post-history

For the creation of the Oprema K&K were awarded the 2[nd] Class National Prize on 7 Oct. 1955.[107] After the completion of the Oprema Kämmerer worked on the design of two further computer models, the ZRA 1 and the ZRA 2. The ZRA 1 was produced in a small series of about 32 machines in the years 1959 to 1963. It was about 40 times faster than one Oprema machine. The ZRA 2 was still in the development stage when the project was abandoned at the end of 1961. Both of them were electronic computers whose logic circuits were based on magnetic cores.[108]

K&K left CZJ at the end of 1959. Kortum became the director of the newly created Central Institute for Automation (ZIA) in Jena and Kämmerer became his deputy. When ZIA in Jena was closed on 31 March 1961 Kämmerer became the director of a newly created satellite facility of the Institute of Mathematics of the Academy of Sciences.[109]

Beginning in autumn 1956 Kämmerer gave a series of courses on "program-controlled computers" and "information theory" as an associate lecturer at the FSU in Jena. On 11 Dec. 1958 he completed the "Habilitation" [a postdoctoral qualification] at the FSU with the thesis "Ziffernrechenautomat mit Programmierung nach mathematischem Formelbild [Digital computer automaton programmed with mathematical formulæ]". On 1 August 1960 he became a Lecturing Professor (Professor mit Lehrauftrag) at the FSU and continued to give courses on computers, cybernetics, information theory and related topics, while his main occupation remained the work at ZIA resp. the Academy of Sciences. These teaching activities ended with his retirement in the summer 1970.[110] In the 1960s and 1970s Kämmerer published several books on computers and cybernetics.[111] In the 1960s K&K, together with Helmut Thiele, were the editors of the German edition of the Soviet publication *Problemy kibernetiki*.[112] Together with Thiele Kämmerer started the Journal "Elektronische Informationsverarbeitung und Kybernetik: EIK [Electronic Information Processing and Cybernetics]" in 1965. In 1970 he was elected to the Leopoldina, one of the oldest existing scientific academies. After reunification

Munich on 23 October 1991 (Fig. 18). In 1994 Konrad Zuse, who also had artistic abilities and who had once pondered to become an advertising artist, created a chalk drawing of Kämmerer. Wilhelm Kämmerer died on 15 August 1994 after he had suffered a heart attack shortly after his 89th birthday.[113]

As mentioned above, Kortum became director of the Central Institute for Automation in Jena on 1 Jan. 1960. When the institute in Jena was closed on 31 March 1961, Kortum became the director of the newly created Forschungsstelle für Meßtechnik und Automatisierung (FMA) der Akademie der Wissenschaften [Research Center for Measurement Technology and Automation of the Academy of Sciences]. On 1 Nov. 1960 Kortum became a Lecturing Professor for Control Engineering and Automation at the Institute of Technology in Ilmenau, while his main occupation remained the work at ZIA resp. FMA. He held this lectureship for the following 10 years. During the 1960s Kortum worked on bolometers, vacuum thermopiles, pyrometers, infrared detectors and similar devices. In 1959 he had become a member of the Central Research Council (Forschungsrat) of the GDR and on 19 Jan. 1960 he became the first president of the newly founded Deutsche Messtechnische Gesellschaft (DMTG) [German Association for Measurement Technology]. In addition to the National Prize in 1955 Kortum received the following awards:

1967 Medal of Merit of the GDR
1971 Merited Technician of the People
1971 Gold Medal at the Leipzig Spring Trade Fair
        for his vacuum thermopile.

Due to a very serious illness Kortum prematurely retired from the FMA in 1971. Herbert Kortum died on 28 September 1979 shortly after his 72nd birthday.[114]

## Is Herbert Kortum a Computer Pioneer?

In his work diary Lenski mentions Kortum several times:
"1954.04.09   new room plan presented to Dr. Kortum … Kortum: the required space area is quite large, could it be done in a different way? …   Endurance tests for the relays should be started immediately, according to his experience the switching of the relays is absolutely reliable.
1954.04.27   Dr. Kortum proposes wire contacts made of silver.
1954.05.06   Dr. Kortum will initiate preliminary talks about the delivery of the relays.
1954.05.13   Meeting at Dr. Kortum's: told to prepare a schedule, a cost estimate and a quantity estimate.
1954.05.14   Drafts of schedule and cost estimate presented to Dr. Kortum.
1954.05.15   Copy of the wiring diagram to Dr. Kortum. Schedule and cost estimate for meeting in Berlin delivered.

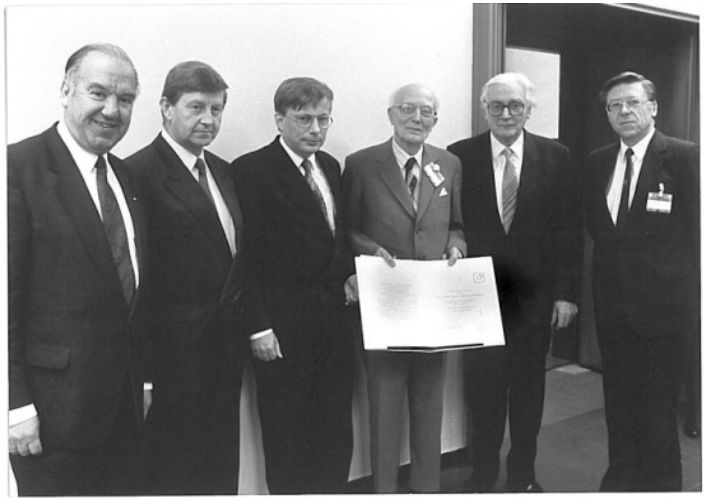

**Figure 18. Kämmerer (3rd from right) receives the Konrad Zuse Medal; Zuse second from right. Courtesy of GI Bonn.**

Kämmerer was awarded the Konrad Zuse Medal of the Gesellschaft für Informatik (German Informatics Society, GI), which was presented to him in

(On 17 May 1954 Kortum presented the Oprema project to minister Rau in Berlin (East), as mentioned above.)

1954.05.18 Meeting at Dr. Kortum's: construction approved. Immediately begin with the work on a large scale.

1954.06.10 Dr. Kortum informed of the current state by telephone.

1954.06.12 Meeting at Dr. Kortum's about the account number, additional personnel and the required space.

1954.06.15 Dr. Kortum is looking for a draftswoman.

1954.08.30 Dr. Kortum did an inspection and asked for a list of the reasons which caused the delay.

1954.09.04 Meeting at Dr. Kortum's: at the visit of the plant manager next week, a plan should be ready which "guarantees" that both machines will be completed on time.

1954.09.10 Dr. Kortum is informed that the whole department ELG M can work for us.

1954.09.11 Meeting at Dr. Kortum's: told to develop a detailed plan by 18 Sep. The plan should have a granularity of 3 days.

1954.09.17 Meeting at Dr. Kortum's: plan is OK.

1954.09.18 (Saturday) Dr. Kortum intends to check the adder circuits on Sunday.

1954.09.29 Dr. Kortum is informed that the estimated costs are about 1.4 million Marks. He says that when the machine is completed nobody will mention this figure.

1954.10.04 (Monday) Dr. Kortum spotted some incorrectly packaged relays on Sunday.

1954.10.11 Dr. Kortum is informed that the milling machine is to be taken away.

1954.11.04 Dr. Schrade, Dr. Kortum and RuMü made a visit."

These entries show that Kortum's role in the Oprema project was essentially a managerial one. This is no surprise, but is explained by the fact that Kortum, as Chief of Development, was responsible for all products of CZJ, from small photographic lenses to the huge planetariums. Seven development labs reported directly to Kortum, among them ELQ, Kämmerer's lab, which was one of the smaller labs.[115]

Nevertheless, there are publications which see Kortum as one of the computer pioneers in Germany, e.g. the 2013 book "Hommage an Konrad Zuse [Homage to Konrad Zuse]" by Friedrich Genser. This claim occurred first in the forerunner publication "German Computer Pioneers" which was published privately (300 copies) by Johannes Jänike and Genser in 1995. It was initiated, financed and edited by Genser and the texts of the biographies were written by Jänike.[116]

Johannes Jänike (1921−2015) studied civil engineering at the Institute for Architecture and Civil Engineering in Weimar in the mid-1950s. During an excursion in 1954 he saw the Oprema, which was then under construction. He later worked as a planning engineer and was also the editor of a series of books on the technology of planning, were he authored a book on computer aided project planning. In 1991, Jänike was awarded the Konrad Zuse Medal of the Zentralverband Deutsches Baugewerbe (ZDB) [Central Association of German Building Industry ].[117]

The German computer pioneer Lehmann (1921−1998), who has already been mentioned in the chapter "Design and Construction" above and who knew Kortum and his work quite well, quite strongly opposed Genser's and Jänike's view in a letter to Jänike dated 19 March 1995, when "German Computer Pioneers" was still in the preparation period. Lehmann refers to the several works by Kortum mentioned throughout the current paper and comes to the conclusion:

"Damit gehört Kortum sicher zu den verdienstvollen Förderern des Einsatzes von Rechentechnik in seinem Industriebereich (also insbes. in den Zeiss-Werken), ein Pionier der Rechentechnik/Informatik wurde er damit jedoch nicht. [Thus, Kortum certainly belongs to the deserving promotors of the use of data processing in his industrial sector (i.e. espec. at the Zeiss works), this, however, did not make him a pioneer of computing/informatics]".

Lehmann developed a series of electronic computers at the Dresden Institute of Technology and was awarded the Konrad Zuse Medal of the Gesellschaft für Informatik (German Informatics Society, GI) in 1989.[118]

Lehmann's view is also corroborated by a publication list published by the Institute of Technology in Ilmenau which for the years 1963 to 1967 cites 25 publications by Kortum on automation, cybernetics, thermo-elements, bolometers and similar items. Two of these papers were co-authored with a second author. One of these two papers is a survey article on "Digitale Informationsmaschinen (Rechenautomaten) [Digital Information Machines (Computing Automata)] co-authored with Fritz Straube, one of the developers of the ZRA 1.[119]

And even Jänike supports Lehmann's view when citing Mühlhausen as follows: "Kortum als Entwicklungshauptleiter ist der Initiator der Rechenmaschinenentwicklung. … Kämmerer ist der theoretische Kopf der Entwicklung. [Kortum, as the Chief of Development, initiated the development of the computers. … Kämmerer was the leading theoretician of the development.]" Lehmann discussed the role of Kortum also in a letter to Jänike dated 23 Dec. 1994 and wrote "Kortum war der Manager [Kortum was the manager]."[120]

Kämmerer himself seemed to be obliged to mention that Kortum had also been involved in the Oprema project when he (Kämmerer) prompted a correctional note in the MTW-Mitteilungen 1957 No.1. In the paper "Die programmgesteuerte …" (see endnote 30) of 1956 Kämmerer had only mentioned himself.[121]

Summing up, we may call Kortum a pioneer computer initiator, as John Ronald Womersley has been called by Brian E. Carpenter and Robert W. Doran.[122]

## Conclusion

Despite the fact that it was technologically a latecomer the Oprema was an important step in the developing of computing at CZJ and the GDR at large. It improved the development of optical systems significantly and paved the way to better lens systems. It also paved the way to the ZRA 1 whose 32 exemplars were running in 15 universities and 15 other institutes and companies GDR-wide during the 1960s. Immo O. Kerner, who was involved in the design of ZRA 1, estimated that more than 15,000 students learnt programming and the basics of computing through this computer, and furthermore, that about 50,000 people had close contact with a ZRA 1 during their education or professional life. The development of Oprema also paved the way to Kämmerer's publications, which have already been mentioned throughout this paper.[123]

Similarly to the Oprema, the current paper itself also seems to be a kind of latecomer, because Jänike and Lehmann discussed in early 1995 the idea to publish the history of the early computers in the GDR in the Annals. On 12 Feb. 1995 Jänike wrote in a letter to Lehmann that he had received a letter from "Mr. Williams, University of Calgary, Alberta, Kanada" accompagnied by many attachments. Michael Roy Williams was at that time professor in the Department of Computer Science at the University of Calgary and Assistant Editor-in-Chief of the IEEE Annals of the History of Computing. Jänike proposed that Lehmann should write about the computer history in Saxony and he would write about the computer history in Thuringia. Lehmann signaled his consent in principle in a letter dated 1 March 1995. Jänike answered on 9 March 1995 and proposed that each of them should draw up a concept for a "Special Issue of the Journal Annals of the History of Computing" on the "History of Computing in East Germany". His concept has the date 16 March 1995 and lists 6 papers whose authors should be he himself, Lehmann or Mühlhausen. However, it seems that this plan did not materialize: none of these papers did appear in the Annals, especially not in the issue 1999, No. 3, which contained a number of papers on early computers in several countries of Eastern Europe. In 2012 a survey paper by James W. Cortada about information technologies in the GDR was published in the Annals, which mentions the Oprema briefly in one paragraph and gives the numbers of relays erroneously as 25,000. As mentioned above, both Oprema machines together contained 16,626 relays.[124]

## Acknowledgments

First of all, I have to thank Georg Elsner, who gave me the decisive nudge to give a talk on the Oprema as my farewell lecture at a symposium at the FSU on 25 October 2008, organized primarily by Michael Fothe. Elsner, a former Zeiss employee, who's now head of his own company, collected a whole bunch of material from Klaus Lösche, a former operator of the Oprema, and from the Zeiss archives. Since I was a complete novice to Zeiss and the Oprema, I could not have given this talk without Elsner's help. In 2014, my colleague Klaus Küspert invited me to give a talk on the Oprema at an alumni event in Jena on 25 October 2014. This then led to the decision to try to write a somewhat more detailed account of the development of this machine. Eberhard Dietzsch, a former lens designer at CZJ, advised me on optical calculations and provided me with Fig. 16 and 17.

For their assistance during my search for information and source material I also thank Helga Kämmerer, Jena, the elder daughter of Wilhelm Kämmerer; Klaus Lösche, Jena; Johannes Jänike (✝), Jena; Wolfgang Koch of Friedrich Schiller University Jena, who gave me one of the Oprema relays which he had inherited from his father; Marte Schwabe and Wolfgang Wimmer of the Zeiss archives, Jena; Margit Hartleb and Rita Seifert of the archives of the Friedrich Schiller University Jena; Wilhelm Füßl and Matthias Röschner of the archives of the Deutsches Museum Munich; Kerstin Weinl of the library of the Technical University Munich; Bernhard Braunecker, emeritus of Leica Geosystems Heerbrugg; Jürg Dedual, the creator of the Virtual Archive of Wild Heerbrugg; Martin Gutknecht, emeritus of the ETH Zürich; Monika Schulte of the Gesellschaft für Informatik, Bonn; Manfred Eyßell, Göttingen; and Horst Zuse, the eldest son of Konrad Zuse.

## About the Author

Jürgen F. H. Winkler is Professor Emeritus at Friedrich Schiller University Jena. During his career at KIT Karlsruhe, Siemens Corporate Research Munich and Friedrich Schiller University Jena he worked and published on programming languages and software engineering. In his farewell lecture in October 2008 he recalled the Oprema and has since then also worked on the history of computers. Read more at http://psc.informatik.uni-jena.de/.

Dresden [On the History of the Institute of Automatic Computing of the Institute of Technology/Technical University Dresden]", p.123, E. Sobeslavsky and N. J. Lehmann, *Zur Geschichte von Rechentechnik und Datenverarbeitung in der DDR 1946–1968* [*On the History of Computing and Data Processing in the GDR 1946–1968*], Hannah-Arendt-Institut, Dresden, 1996, pp. 123–157; H. H. Goldstine and A. Goldstine, "The Electronic Numerical Integrator and Computer (ENIAC)", *Mathematical Tables and Other Aids to Computation (MTAC, 0891-6837)*, vol. 2, no. 15, July 1946, pp. 97-110, Lehmann erroneously cites this as MTAC vol. 2, no. 13 (April 1946), Lehmann, "Zur Geschichte des »Instituts für maschinelle Rechentechnik«", p. 154.

29 G. Lenski, "Arbeitsberichte [Work Reports]", UACZ, BACZ 27995; E. Mühlhausen, "Am Anfang war OPREMA . . . [In the Beginning was OPREMA . . .]", *rechentechnik/datenverarbeitung* (0300-3450), vol. 24, no. 7, 1987, pp. 34–36; Jänike and Kämmerer, "Die Insel des Vergessens", p. 154.

30 LWD, 6 and 8 April 1954; W. Kämmerer and H. Kortum, "Oprema, die programmgesteuerte Zwillings-Rechenanlage des VEB Carl Zeiss Jena"; W. Kämmerer, "Die programmgesteuerte Rechenanlage „Oprema" [The program-controlled computing device „Oprema"]", *MTW-Mitteilungen*, vol. III, no. 5, 1956, pp. 127-132; W. Kämmerer, *Ziffernrechenautomaten* [*Digital Computing Automata*], Akademie-Verlag, Berlin, 1960, p. 121; IEEE Std 754™-2008, sect. 2.1.49; ISO/IEC/IEEE 60559: 2011(en), sect. 2.1.49; D. E. Knuth, *Seminumerical Algoritms*, 2nd ed., Addison-Wesley, Reading et al, 1981, 0-201-03822-6, pp. 198–199.

31 R. (?) Dietrich, Memo 29. 3. 1955, UACZ, BACZ No. 19570.

32 Ministerium für Maschinenbau [Ministry of Machine Building], Sekretariat des Ministers [Office of the Minister], Memo 19 May 1954; H. Kortum, "Trip Report", 20 May 1954; H. Kortum, "Kurze Zusammenfassung des Vortrages zur Oprema, gehalten vor Min. Rau am 17.5.54 [Short summary of the presentation given to minister Rau on 17 May 1954]", p. 1; all in UACZ, BACZ 23789.

33 J.F.H. Winkler, "Konrad Zuse und die Optik-Rechenmaschine in Jena [Konrad Zuse and the Optic Computer at Jena]", *Log In* (0720-8642), no. 166/167, 2010, pp. 132–136; J.F.H. Winkler, "Konrad Zuse and Switzerland", Swiss Physical Society, *SPS Communications*, no. 34, 2011, pp. 42–44; for the slides of two talks about the Oprema see http://psc.informatik.uni-jena.de/hist/oprema.htm , retr. 2019Apr05.

34 "DM" is here the currency of the GDR and not the Deutschmark which was usually also referred to by the acronym "DM". From 24 June 1948 until 31 July 1964 the official name of the GDR-Mark was "Deutsche Mark der Deutschen Notenbank (DM) [German Mark of the German Treasury (DM)]", Bundesministerium für gesamtdeutsche Fragen (Hrsg.), *SBZ von A bis Z*, Bonn, 1966, Deutscher Bundes-Verlag, 10. Aufl., p. 519 [Federal Department for All-German Affairs, ed., *SBZ from A to Z*, German Federal Publishing House, Bonn, 10th ed., 1966].

35 T. P. Hughes, "ENIAC: Invention of a Computer", *Technikgeschichte* (0040-117X), vol. 42, no. 2, 1975, pp. 148–165, here pp. 156–157.

36 H. Thiele, *Carl Zeiss Jena – Entwicklung und Beschreibung der Photoobjektive und ihre Erfinder* [*Carl Zeiss Jena – Development and Description of the Photographic Lenses and their Inventors*], 2nd ed., privately published, München, 2007, p. 37; telephone conversation between E. Dietzsch and the author, 14 May 2018.

37 Invitation to and other material about the Oprema Party, UACZ, BACZ 27995.

38 MDR (Mitteldeutscher Rundfunk [Broadcasting Service of Central Germany]), "Einblicke – 150 Jahre Carl Zeiss [Insights – 150 Years Carl Zeiss]", 13 Nov. 1996, 20:15 (30 minutes), here position 17:45; Invitation to and other material about the Oprema Party.

39 K. Schumann, "Begründung für einen Antrag auf Auszeichnung der wissenschaftlichen Mitarbeiter des VEB Carl Zeiss Jena, Dr. Herbert Kortum und Dr. Wilhelm Kämmerer, mit dem Nationalpreis [Rationale for a petition to award the National Prize to the scientists Dr. Herbert Kortum and Dr. Wilhelm Kämmerer of VEB Carl Zeiss Jena]", 13 April 1955, p. 4, UACZ, BACZ 19570; W. Kämmerer, "Ausführlicher Abschlußbericht – Feinjustierung der Oprema [Detailed and final report – Vernier adjustment of the Oprema], 15 Jan 1956, p. 4, UACZ, BACZ 21563.

40 UACZ BI 02448, 5 Jan. 1955; see also endnote 37; E. Mühlhausen, "OPREMA und ZRA 1 – Frühe Entwicklungen der digitalen Rechentechnik im Zeisswerk Jena [OPREMA and ZRA 1 – Early Developments of Digital Computing in the Zeiss Works Jena]", Kramer, Lothar, ed., *Jenaer Jahrbuch zur Technik- und Industriegeschichte Band 2* [*Jenaer Yearbook on the History of Technology and Industry*, vol. 2], Glaux-Verlag, Jena, 1999, 3-931743-10-1, S. 109–127, here pp. 112–113; the Oprema people may have known the Cornel-Trio version of the song "Bravo, bravo, beinah wie Caruso [Bravo, bravo, almost like Caruso]",